\documentclass[
aps,prd,reprint,amsmath,
 nofootinbib,
]{revtex4-2}

\usepackage{hyperref}
\usepackage{dcolumn}
\usepackage{comment}
\usepackage{graphicx}
\usepackage{txfonts}
\preprint{APS/123-QED}

\usepackage{xcolor}
\usepackage{multirow}
\usepackage{array}
\usepackage{ulem}
\usepackage{soul} 
\usepackage{xspace}
\usepackage{subcaption}

\usepackage{ragged2e}

\usepackage{tikz}
\usetikzlibrary{shapes.geometric, arrows.meta, positioning}
\tikzstyle{process} = [rectangle, draw=black, text centered, minimum height=1cm]
\tikzstyle{ligerbox} = [rectangle, draw=black, fill=gray!10, text centered, minimum height=1cm]
\tikzstyle{arrow} = [thick,->,>=stealth]
\tikzstyle{rounded} = [rectangle, draw=black, text centered, minimum height=1cm, rounded corners=6pt]
\tikzstyle{redproc} = [rectangle, draw=red, text centered, minimum height=1cm]
\tikzstyle{redliger} = [rectangle, draw=red, fill=gray!10, text centered, minimum height=1cm]
\tikzstyle{redarrow} = [thick,->,>=stealth, draw=red]
\tikzstyle{redrounded} = [rectangle, draw=red, text centered, minimum height=1cm, rounded corners=6pt]
\tikzstyle{titlebox} = [rectangle, draw=none, text centered, font=\large\bfseries]

\RequirePackage{commandsMYK}
\begin{document} 

\title{The impact of our peculiar motion on primordial non-Gaussianity measurements using the \texttt{LIGER4GAL} framework}

\author{Bartolomeo Bottazzi Baldi}
 \email{bartolomeo.bottazzibaldi@phd.unipd.it}
 \affiliation{
 Dipartimento di Fisica e Astronomia Galileo Galilei, Università di Padova, 35131, Padova, Italy
 }
 \affiliation{
 INFN Sezione di Padova, I-35131, Padova, Italy
 }
\author{Mohamed Yousry Elkhashab }%
\affiliation{%
 Dipartimento di Fisica – Sezione di Astronomia, Università di Trieste, Via Tiepolo 11, 34131
    Trieste, Italy
}%
\affiliation{
INAF, Osservatorio Astronomico di Trieste, Via Tiepolo 11, I-34131 Trieste, Italy  
}
\affiliation{
INFN, Sezione di Trieste, Via Valerio 2, 34127 Trieste TS, Italy
}
\affiliation{
IFPU, Institute for Fundamental Physics of the Universe, via Beirut 2, 34151 Trieste, Italy
}
\author{Daniele Bertacca}
\affiliation{
 Dipartimento di Fisica e Astronomia Galileo Galilei, Università di Padova, 35131, Padova, Italy
 }
 \affiliation{
 INFN Sezione di Padova, I-35131, Padova, Italy
 }
 \affiliation{INAF, Osservatorio Astronomico di Padova, Italy}
 \author{Cristiano Porciani}
\affiliation{Argelander-Institut für Astronomie, Auf dem Hügel 71, D-53121 Bonn, Germany}
 
  \begin{abstract}
  Current and forthcoming galaxy surveys will map the observable Universe with unprecedented depth, sky coverage, and precision. These maps are affected by relativistic redshift-space distortions (RSDs), which become increasingly relevant on ultra-large scales. Accurate modelling of these relativistic RSDs is essential to avoid systematic biases in key cosmological measurements, such as primordial non-Gaussianity (PNG).  To address this, we introduce an updated implementation of the \texttt{LIGER} method, \texttt{LIGER4GAL}, which incorporates all linear-order relativistic RSDs directly at the tracer level of high-resolution N-body simulations. We demonstrate that \texttt{LIGER4GAL} improves upon previous iterations of the \texttt{LIGER} method by reproducing the expected non-linear clustering while maintaining accuracy for relativistic RSDs on large scales. We use the updated code to generate a DESI-like sample of luminous red galaxies from the Huge MultiDark Planck simulation.
  By measuring the power spectrum multipoles of this sample with and without the imprint of relativistic RSDs, we assess the impact of relativistic effects on measurements of the local PNG signal ($f_\mathrm{nl}$). 
  We find that the omission of the “finger-of-the-observer” (sourced by the peculiar velocity of the observer) effect in the power spectrum modelling can bias measurements of $f_{\rm nl}$ by more than $1$ ($0.25$) $ \sigma_{f_{\rm nl}}$ in 40\% (80\%) of the possible realizations of the universe if scales down to $k_\mathrm{min} = 0.0015\,h/\mathrm{Mpc}$ are included.
  \end{abstract}
  
\maketitle
%
\vspace*{-1.4cm}
\tableofcontents

\section{Introduction}
Forthcoming and ongoing surveys such as DESI \citep[][]{desicollaboration2016desiexperimentisciencetargeting}, Euclid \citep[][]{Euclid_Over_2025}, and SPHEREx \citep[][]{doré2015cosmologyspherexallskyspectral} are designed to map the large-scale structure of the Universe in unprecedented volume, depth, and precision. These missions aim to advance our understanding of fundamental physics by probing the nature of Dark Energy (DE) and Dark Matter (DM), constraining the initial conditions of the Universe, and testing the validity of general relativity on cosmological scales. 

Typically in galaxy redshift surveys, we assume a homogenous and isotropic universe, modelled by a Friedmann-Lema\^{\i}tre-Robertson-Walker (FLRW) metric, to convert from redshifts to distances. However, the inhomogeneities of the observed Universe arising from structure formation introduce additional contributions to the observed redshift beyond the purely cosmological component, which are collectively referred to as redshift-space distortions (RSDs). 
RSDs are typically divided into two categories. The first comprises local effects, such as the leading-order contribution from the peculiar velocities of source and observer galaxies, first identified by \citet{1987MNRAS.227....1K}, as well as distortions arising from the local gravitational potential at both the emitter and the observer \citep[][]{Alam:2017izi}. 
The second category concerns line-of-sight integrated effects, such as weak lensing \citep{Sasaki:1988vk,Matsubara_2000,matsubara_constraining_2001}, which is the dominant integral contribution. Furthermore, there are other integrated terms such as the integrated Sachs-Wolfe effect \citep{Sachs-Wolfe67} and the Shapiro time delay \citep{Shapiro_time}. We collectively refer to all contributions beyond the source peculiar velocity term (which is routinely included in cosmological analyses) as relativistic RSDs. 
The impact of relativistic RSDs on the galaxy overdensity has been investigated both at linear order \citep[e.g.][]{PhysRevD.80.083514, PhysRevD.84.063505, PhysRevD.84.043516, PhysRevD.85.023504,Bertacca_2012} and at second order \citep[e.g.][]{Yoo_2014,Dio_2014,Bertacca_2015,bertacca_2nd_mag,Umeh_2017}. In particular, several studies have quantified their effect on observables such as the multipoles of the power spectrum \citep{Elkhashab_2021,castorina_observed_2022,noorikuhani_wide-angle_2023} and the two-point correlation function \citep{raccanelli_large-scale_2014,Bertacca_2015,Tansella_2018,breton_imprints_2019,Elkhashab_2025,Lepori_2025}. 

The main aim of this study is the quantitative analysis of systematic biases in key cosmological measurements that arise from neglecting the impact of relativistic RSDs. Firstly, we focus on measurements of primordial non-Gaussianity (PNG) derived from large-scale-structure (LSS) analyses. PNG serves as a pivotal probe of the physics of the early Universe, offering a direct window into the inflationary mechanisms that generate the primordial fluctuations, which in turn seed structure formation \citep{Bartolo_2004,  alvarez2014testinginflationlargescale}.
In the local form (typically described by the non-Gaussianity parameter $f_{\rm nl}$), PNG induces a scale-dependent correction to the bias of tracers of the underlying DM field, which becomes most prominent on large scales \citep{Dalal_2008, Matarrese_2008,Slosar_2008,Desjacques_2010, 2010PhRvD..81f3530G}.
However, relativistic effects impact the observed power-spectrum signal in this same scale regime, where some of the contributions follow a similar scale dependence as the PNG signature \citep[e.g.][]{bruni_disentangling_2012,camera_einsteins_2015,Raccanelli_2016,Foglieni+23}. Ignoring  these effects can thus lead to systematic biases in $f_{\rm nl}$ measurements   \citep[e.g.][]{Bruni:2011ta,Bahr_Kalus_2021,guedezounme2024primordialnongaussianityeffects}. 

Secondly, we focus on constraints of the growth rate of structure obtained through standard RSD analyses \citep{Peacock_2001}. These studies extract cosmological information by analysing the n-point statistics of tracers using perturbative models that incorporate structure formation, biasing, and velocity-induced RSDs up to linear and quasi-linear scales. One widely adopted approach is the Effective Field Theory of LSS \citep[EFT, for a review, see][]{ivanov2022effectivefieldtheorylarge}, which has been applied to datasets such as the BOSS survey \citep{d_Amico_2020} and, more recently, in the DESI collaboration \citep{maus2024comparisoneffectivefieldtheory}. However, in its standard formulation, EFT-based RSD modelling accounts only for the velocity contribution, neglecting relativistic RSDs. In this work, we examine the validity of that assumption and quantify the potential impact of missing contributions, focusing on constraints of the power spectrum amplitude ($A_{\rm s}$), the DM density parameter ($\omega_{\rm c}$) and the dimensionless Hubble parameters ($h$) within a vanilla $\Lambda$CDM model.  

To measure the impact of relativistic RSDs on these analyses, we generate mock galaxy catalogues that include relativistic RSDs up to linear order using the \liger method and use these catalogues for cosmological analysis. To that end, we implement a new version of the method, hereafter \ligerGAL, designed to generate lightcones with improved accuracy at non-linear scales of the matter overdensity while maintaining the linear implementation of relativistic RSDs.  

Previous iterations of the \liger method \citep{Borzyszkowski:2017ayl,Elkhashab_2021,Elkhashab_2025} relied on large, low-resolution simulations and employed a simple linear biasing scheme to construct galaxy mocks tailored to specific surveys. While this approach was effective at reproducing the clustering statistics on large scales, it failed in describing the clustering on smaller scales due to the resolution of the underlying simulations, and to the biasing scheme implemented.
In contrast, \ligerGAL is applied directly to halo catalogues, enabling the use of high-resolution simulations and significantly extending the validity of the resulting mock catalogues into the non-linear regime, which allows us to perform realistic cosmological analysis of the dataset.
To populate haloes with galaxies, we also provide a toolkit (which is compatible with the \ttt{halotools} framework \cite{Hearin_2017}) that ensures consistency with the relativistic treatment of \ligerGAL, by imprinting the relativistic corrections on each galaxy's position and magnification, accounting also for their peculiar velocities with respect to the host halo.

In this work, we apply the new code to one realization from the MultiDark  simulation suite \citep[][]{2013AN....334..691R, 2012MNRAS.423.3018P, 2016MNRAS.457.4340K}, to generate galaxy catalogues that follow the expected distribution of DESI luminous red galaxies (LRGs) in both full-sky and partial-coverage configurations. Specifically, we make use of the Huge MultiDark Planck (HMDPL) simulation, which is  a DM-only simulation with a box size of $4\,\mathrm{Gpc}/h$ and a mass resolution of $m = 7.9 \times 10^{10} \, \mathrm{M}_\odot/h$. These catalogues allow us to assess the impact of not accounting for relativistic RSDs on inferred cosmological parameters in two contexts: \textit{(i)} full-shape fits of the power spectrum multipoles using the EFT model,
and \textit{(ii)} PNG analyses of the power spectrum multipoles. 

This paper is structured as follows. In Sect. \ref{theory}, we first review the theoretical framework for relativistic RSDs and their implementation in \liger, then introduce \ligerGAL, highlighting its improvements over the earlier low-resolution version. In Sect. \ref{Methods}, we describe the \ligerGAL approach, and its application to the HMDPL simulation, producing DESI-like LRG catalogues. Sect. \ref{Results} presents the validation of \ligerGAL against theoretical predictions and the previous version of the \liger method (hereafter \ligerDM), followed by an assessment of the impact of relativistic effects—particularly the FOTO signal—on the inference of various cosmological parameters. We summarize and conclude in Sect.~\ref{sec:conc}.

Throughout this paper, we adopt Einstein's summation convention and  define the space-time metric tensor to have the  
signature $(-,+,+,+)$. Greek indices refer to space-time components (i.e. run from 0 to 3), while Latin indices label spatial components (i.e. run from 1 to 3).  Furthermore, the Dirac delta and 
the Kronecker delta functions are denoted by the symbols $\delta^{\rm D}$ and $\delta^{\rm K}$, respectively. Our Fourier-transform convention is $\tilde{f}(\bs{k}) = \int f(\bs{x})\,\mathrm{e}^{-i\bs{k}\cdot \bs{x}}\,\dif ^3 x $, and $f(\bs{x}) = 1/(2\pi)^3\int \tilde{f}(\bs{k})\,\mathrm{e}^{+i\bs{k}\cdot \bs{x}}\,\dif ^3 k $.  
Finally, the symbol $c$ denotes the speed of light in vacuum. 
\section{The \liger method}
\label{theory}
\subsection{Relativistic redshift space distortions}

Assuming a homogeneous and isotropic Friedmann–Lemaître–Robertson–Walker (FLRW) universe, we can use the redshift measured in galaxy surveys, $z_{\rm obs}$, as a proxy for the comoving distance. This relation can be written as  
\begin{equation}
\label{eq:com_dist}
x = \int^{z_{\rm obs}}_0 \frac{c}{H(z)}\,\dif z\,,
\end{equation}
where $H(z)$ denotes the Hubble parameter. Combined with the observed angular positions on the sky, $\bs{n}\equiv \bs{x}/x$, we are able to construct maps of the Universe. Such maps, which are model dependent due to Eq.~\eqref{eq:com_dist}, are then used for various cosmological statistical studies.

However, Eq.~\eqref{eq:com_dist} neglects the impact of inhomogeneities encountered by light along its trajectory from source to observer. These inhomogeneities alter the apparent position of the source, as first demonstrated in the seminal work of \citet{1987MNRAS.227....1K}. Using  linear perturbation theory, Kaiser has shown that the leading-order correction on large scales -- arising from the peculiar velocities of the source galaxies -- induces a line-of-sight-dependent distortion to the observed galaxy overdensity $\delta_{\rm g}$, with respect to the (unobservable) real-space galaxy overdensity $\delta_{\rm g,r}$.
Nevertheless, the fact that we observe galaxies on our past light cone implies that the photons are not influenced only by local peculiar velocities but also by the entire distribution of matter they encounter. Relativistic linear treatments  have shown that additional effects—such as the integrated Sachs–Wolfe effect, weak gravitational lensing, and Shapiro time delay—also affect $\delta_{\rm g}$. These corrections typically become significant at large scales \citep[e.g., see ][]{PhysRevD.80.083514, PhysRevD.84.063505, PhysRevD.84.043516, PhysRevD.85.023504}. The numerical modelling of these effects, collectively called relativistic RSDs, is the object of this study. We summarize the linear theoretical treatment of these effects in this section. We limit the implementation of relativistic corrections at linear order, as we are mainly interested in their impact on the two-point statistics of galaxy clustering, where linear theory suffices.

To explicitly quantify the impact of relativistic RSDs, a specific gauge choice is required. In this work, we adopt the scalar-restricted Poisson gauge, i.e. 
\begin{equation}
\label{eq:metric}
\dif s^2 = a^2(\eta)\left[-(1+2\Psi)\,c^2\,\dif\eta ^2 +(1-2{\Phi})\,\delta^{\rm K}_{ij}\, \dif x^i\;\dif x^j\;\right]\,,
\end{equation}
where $\eta$ is the conformal time, $a$ is the scale factor, the two potentials $\Phi$ and $\Psi$ are the (dimensionless) Bardeen potentials \citep{Bardeen:1980kt}.  Using the perturbed geodesic equation, it is possible to define a coordinate map that relates the observed four-vector, $x^\mu_s$, to the real-space four-vector, $x^\mu_r$. The  coordinate map  $\Delta x^\mu=x^\mu_r-x^\mu_s$, is given by  \citep{PhysRevD.80.083514, PhysRevD.84.063505, PhysRevD.84.043516, PhysRevD.85.023504}
\begin{equation}
\begin{split}
\Delta x^{0} =& \frac{c}{\mc{H}}\,\delta \ln a\,, \\
\Delta x^{i} =& -x\left[ n^i_{\rm{s}}\,(\Phi_{\rm{o}}+\Psi_{\rm{o}} )+ {\varv^i_{\rm{o}}\over c} -{n^i_{\rm{s}}\;(n^j_{\rm{s}}\varv_j)_{\rm{o}}\over c}  \right] - \frac{c\,n^i_{\rm{s}}}{\mc{H}} \delta \ln a \\ 
& + \int^{x}_0 {(x - x')\over c}\left[n^i_{\rm{s}}\partial_0(\Phi+\Psi)-\delta^{i}_j\partial^j(\Phi+\Psi)\right]\dif x'\\
&+ 2 n^i_{\rm{s}} \int^{x}_0 (\Phi+\Psi) \,\dif x'\,,
\label{eq:shift_lig}
\end{split}
\end{equation}
where $\partial_0=\partial/\partial\eta$, $\partial_i=\partial/\partial x^i$, and   
\begin{equation}
 \delta \ln a = \left[{\frac{({\varv}^{i}_\mathrm{e}-{\varv}^{i}_\mathrm{o})}{c} }\cdot {{n_{\rm{s},i}}} - (\Phi_\mathrm{e}-\Phi_\mathrm{o}) - \int_0^x \frac{\partial^0(\Phi + \Psi)}{c}\;\dif x'\right]\,.
 \label{eq:deltalna}
\end{equation}
Here $\varv^i$  denotes the peculiar velocity vector, $\mc{H}=\partial_0 \mathrm{ln}a$ is the conformal Hubble parameter  and the subscripts ``o" and ``e" refer to quantities evaluated at the observer's and emitter's positions, respectively.
\footnote{Equation ~\ref{eq:deltalna} ignores corrections at the observer's location of the scale factor $a$,
 as they only impact the mean number density and have
no effect on the observed overdensity \citep{Bertacca_2020}.}
 
Cosmological perturbations also alter the solid angle under which galaxies are seen by distant observers, thus enhancing, or decreasing their apparent flux \citep[e.g.][]{Broadhurst_1995}. 
In terms of the luminosity distance, $d_{\rm L}$, the magnification of a galaxy is defined as
\begin{equation} 
 {\mathcal M}=\left( \frac{d_{\rm L}}{\bar{d}_{\rm L}}\right)^{-2}\;, 
\end{equation} 
where $\bar{d}_{\rm L}$ denotes the luminosity distance in the background model universe evaluated at $z_\mathrm{obs}$.
At linear order  \citep[e.g.][]{PhysRevD.84.043516, Bertacca_2015}, 
\begin{equation}
\label{eq:Mag_lig}
\begin{split}
\mc{M}(\bs{x}) =&  1 + 2\Phi_{\rm e} - \,2\left(1-\frac{c}{\cH x} \right)\,\delta \ln a - 2\,\frac{\varv^{i}_{o}}{c}\cdot {n_{\rm{s},i}} \\
&+ {2\,\kappa} - \frac{2}{x}\int^{x}_0 (\Phi+\Psi)\; \dif x'\,\,.
\end{split}
\end{equation}
where  $\kappa$ is the weak-lensing convergence
\begin{equation}
    \kappa(\bs{x}) = \frac{1}{2}\int_0^x (x-x')\frac{x'}{x}\nabla_\perp^2(\Phi+\Psi)\,\mathrm{d} x'\,.
\end{equation}
Equipped with the distortions in the observed position and flux, we are able to derive how the observed galaxy overdensity of a catalogue of a particular tracer is affected. To that end, let us define the average comoving density assuming flux-limited tracer
 \begin{equation}
    \bar{n}_{\mathrm{g}}(z) = \bar{n}\left(L_\mathrm{lim}(z),z\right)\,,
\end{equation}
where  $L_\mathrm{lim}(z)$ is the luminosity limit associated with a flux cut $f_\mathrm{cut}$ at a given redshift $z$. 

Following the definitions in \cite{Bertacca_2015,Elkhashab_2021}, we can then quantify  the rate of change with redshift and the luminosity limit of the survey via the evolution
\begin{equation}
    \mathcal{E}(z) = - \frac{\partial\,\mathrm{ln}\,\bar{n}(L_\mathrm{min},z)}{\partial\,\mathrm{ln}\,(z+1)}\Bigg|_{L_\mathrm{min}=L_\mathrm{min}(z)}\,
\end{equation}
and magnification
\begin{equation}
    \mathcal{Q}(z) = - \frac{\partial\,\mathrm{ln}\,\bar{n}(L_\mathrm{min},z)}{\partial\,\mathrm{ln}\,L_\mathrm{min}}\Bigg|_{L_\mathrm{min}=L_\mathrm{min}(z)}
\end{equation}
 biases, respectively. Putting all the distortions together, the impact on the observed overdensity can be written as \citep{PhysRevD.80.083514, PhysRevD.84.063505, PhysRevD.84.043516, PhysRevD.85.023504}
\begin{equation}
\label{eq:Deltag}
\begin{split}
\delta_{\rm g} (\bs{x})
 =&  \delta_{\rm g, r}-\frac{1}{\cH}\frac{\partial (\bs{\varv}_\mathrm{e}\cdot \bs{n})}{\partial x}- 2\,(1-\mc{Q}) \,{\kappa}\\
 &+\left[2- \mc{E}  + \frac{\partial_0 \cH}{\cH^2} + \frac{2\,(1-\mc{Q})\,c}{x\, \cH}\right]\frac{\bs{\varv}_\mathrm{o}}{c}\cdot \bs{n}\\
    &+
 \left[ \mc{E}-2\mc{Q}  - \frac{\partial_0 \cH}{\cH^2} - \frac{2(1-\mc{Q})\,c}{x\, \cH}\right]\\&%
\times\left[\frac{\bs{\varv}_\mathrm{e}}{c}\cdot \bs{n} - (\Phi_\mathrm{e}-\Phi_\mathrm{o}) - \int_0^x {\frac{\partial^0(\Phi + \Psi)}{c}}\;\dif x'\right]
 \\
 & -2\,(1-\mc{Q})\,\Phi_\mathrm{e}+\Psi_{\rm e} + {\frac{\partial^0 \Phi_{\rm e}}{\mc{H}}} +  \left(3- \mc{E} \right)\frac{\mc{H}\velpot}{{c}^2} \\
 & +\frac{2\,(1-\mc{Q})}{x} \int_0^x(\Phi+\Psi)\;\dif x'\,,
 \end{split}
\end{equation}
 where we denote the linear velocity potential by $\velpot$.
Equation~\eqref{eq:Deltag} analytically describes the impact of relativistic RSDs on the observed overdensity. The second term denotes  the standard Kaiser correction and represents the leading-order contribution. The next-to-leading order is the weak lensing term \citep[e.g.][]{Sasaki:1988vk,Matsubara_2000}, which is particularly significant in photometric surveys \citep{EUCLID_PHOTOMETRIC_WP9}. Of particular relevance to this study is the subsequent term, sourced by  peculiar velocity of the observer. This contribution has received increasing attention due to its potential to be isolated and used as a probe of the observer’s motion
\citep{Bertacca_2020,Bahr_Kalus_2021,Elkhashab_2025}, thereby offering insight into the apparent tension between the kinematic dipole inferred from early-Universe CMB measurements \citep{2020}, and the one measured from late-Universe quasar and radio catalogues \citep[e.g.,][]{Secrest_2021,Bohme:2025nvu}. The remaining terms are proportional to the gravitational potential and are sub-leading in the standard clustering statistics \citep[i.e, the power spectrum and two-point-correlation function][]{euclidcollaboration2024euclidpreparationimpactrelativistic}. However, several studies have explored the possibility of isolating these contributions, including the Doppler velocity term \citep{Borzyszkowski:2017ayl,Montano_2024}.

\subsection{\liger}

\begin{figure*}[ht]
\centering
\begin{tikzpicture}[node distance=1.5cm and 2cm]

\node (nbodyL) [process] {Low-res N-body};
\node (particles) [rounded, below of=nbodyL] {DM particles};
\node (coordL) [rounded, below of=particles] {\begin{minipage}{3.5cm}
\centering
\liger \\ Coordinate transform
\end{minipage}};
\node (fields) [rounded, below of=coordL] {Matter fields};
\node (biasing) [rounded, below of=fields] {Biasing};
\node (survey) [ligerbox] at ([xshift=3.0cm] biasing.east) {
  \begin{minipage}{3.5cm}
  \centering
  \textbf{Survey functions:}\\
  $\bar{n}_{\mathrm{g}}(z),\,b(z),\,\mathcal{E}(z),\,\mathcal{Q}(z)$
  \end{minipage}
};
\node (lightconeL) [process, below of=biasing] {Galaxy catalogue};

\node (nbodyR) [redproc, right=6cm of nbodyL] {N-body};
\node (tracers) [redrounded, below of=nbodyR] {Halo merger tree};
\node (coordR) [redrounded, below of=tracers] {
\begin{minipage}{3.5cm}
\centering
\ligerGAL \\ Coordinate transform
\end{minipage}
};
\node (lightconeR1) [redrounded, below of=coordR] {Halo catalogue};
\node (HOD) [redrounded, below of=lightconeR1] {HOD};
\node (hodpars) [redliger] at ([xshift=3.5cm] HOD.east) {HOD parameters};
\node (lightconeR2) [redproc, below of=HOD] {Galaxy catalogue};

\draw [arrow] (nbodyL) -- (particles);
\draw [arrow] (particles) -- (coordL);
\draw [arrow] (coordL) -- (fields);
\draw [arrow] (fields) -- (biasing);
\draw [arrow] (survey) -- (biasing);
\draw [arrow] (biasing) -- (lightconeL);

\draw [redarrow] (nbodyR) -- (tracers);
\draw [redarrow] (tracers) -- (coordR);
\draw [redarrow] (coordR) -- (lightconeR1);
\draw [redarrow] (lightconeR1) -- (HOD);
\draw [redarrow] (hodpars) -- (HOD);
\draw [redarrow] (HOD) -- (lightconeR2);

\node[titlebox, above=of nbodyL, yshift=-0.5cm] {\ligerDM};
\node[titlebox, above=of nbodyR, yshift=-0.5cm] {\ligerGAL};

\end{tikzpicture}
\caption{\justifying \textit{Left}: Field approach schematic of \ligerDM. \textit{Right}: \ligerGAL approach implemented in this work. }
\label{fig:liger_flowchart}
\end{figure*}
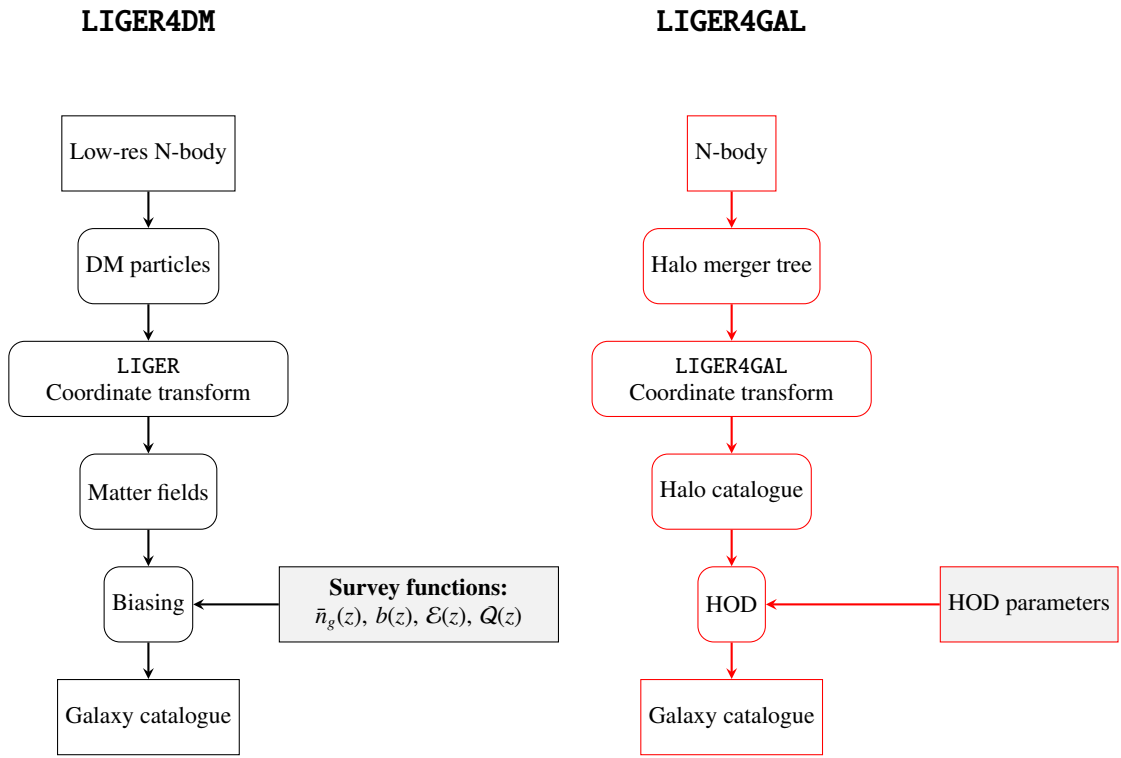

The \liger \citep[LIght cones with GEneral Relativity,][]{Borzyszkowski:2017ayl,Elkhashab_2021,2025} method is a numerical technique for generating galaxy catalogues of the Universe on the past light cone of an observer with a $\Lambda \mathrm{CDM}$ cosmology. The method uses the coordinate transformation between real and redshift space (see Eqs.~\ref{eq:shift_lig}) and the magnification (see Eq.~\ref{eq:Mag_lig}) to calculate the observed positions and fluxes of tracers in Newtonian N-body simulations. In this section, we present an updated version of the \liger method for generating light cones from high-resolution simulations after briefly reviewing the low-resolution implementation of \liger employed in previous studies (hereafter \ligerDM).

\subsubsection{\ligerDM}
\label{lowres}

In this section, we summarise the field-based methodology introduced in \citet{Elkhashab_2021,Elkhashab_2025}. This approach was designed to generate galaxy mocks that span very large volumes while keeping computational costs low, yet still preserving the accuracy of large-scale clustering statistics.
\ligerDM applies the coordinate map to the DM particles of a given simulation. For each particle, it computes its perturbed position, magnification as well as the change in redshift (see Eq.~\ref{eq:deltalna}). The accompanying \texttt{buildcone}   toolkit then uses a cloud-in-cell algorithm to construct the matter density contrasts $\delta_{\rm DM, r}(\bs{x},z)$ and $\delta_{\rm DM}(\bs{x},z)$ out of the real and redshift-space positions, respectively. At linear order, we can also construct field-level functions for $\mc{M}$ and $\delta \ln a$ mass-weighted averages \citep[see appendix A2 in][]{Elkhashab_2021}. Equipped with these four functions, we create  a galaxy density field using
\footnote{With a slight abuse of notation, we equate the composite functions $b_1(z(x)),\,\mc{Q}(z(x)) \,\,\& \,\,\mc{E}(z(x))$ and the functions $b_1(x),\,\mc{Q}(x) \,\&\, \mc{E}(x)$.}
\begin{eqnarray}
    n_{\rm g}\left(\bs{x}\right) \approx& \bar{n}_{\rm g}(x)\,\left[\left(b_1(x)-1\right)\,\delta_{\rm DM, r}(\bs{x})+\delta_{\rm DM}(\bs{x})\right.\nonumber\\
    &+\left.\mathcal{E}(x)\,\delta\, \mathrm{ln}\,a(\bs{x})+\mathcal{Q}(x)\,\left(\mathcal{M}(\bs{x})-1\right)\right]\,,
    \label{eq:delta_g_bc}
\end{eqnarray}  
which can be directly derived from  Eq.~\eqref{eq:Deltag} by replacing the real-space galaxy overdensity by the product of the $\delta_{\rm DM, r}(\bs{x})$  and the linear bias  $b_1(x)$.  Equation~\eqref{eq:delta_g_bc} permits the use of the same DM light cone to produce multiple galaxy maps  tailor-made for distinct surveys by changing the functions $\bar{n}_{\rm g}, b,\,\mc{Q} \,\,\text{and} \,\,\mc{E}$, which we denote as ``survey functions". A flowchart of this procedure is shown on the left side of Fig.  \ref{fig:liger_flowchart}.

\subsubsection{\ligerGAL}
\label{liger4gal}

The approach described in Sect. \ref{lowres} performs well for studies at large-scale structures. However, by design it is less accurate in describing the clustering signal on quasi-linear scales. This limitation arises from the use of the linear relation outlined in Eq.~(\ref{eq:delta_g_bc}), which neglects the non-linear contributions that arise in the matter-tracer bias relation. 

To circumvent this issue, in this work we apply the \textsc{LIGER} method directly to biased tracers of the DM field, without relying on Eq.~\eqref{eq:delta_g_bc}, which integrates easily with our numerical framework and facilitates the implementation of relativistic effects in standard analyses, where significantly more information about tracers is retained compared to DM.
We thus present a new version  of the  \liger method (\ligerGAL)  that directly post-processes large, high-resolution Newtonian simulations, avoiding the need to rely on the biasing procedure. 
By shifting directly the tracer under study (instead of the simulation particles), we can generate a catalogue that is consistent with both the large-scale signature arising from relativistic RSDs, and with the clustering signal of the input simulation up to non-linear scales. 

The general workflow of \ligerGAL is as follows. Starting from a halo catalogue, we reconstruct the corresponding halo merger tree (see appendix~\ref{lightcone}) and the gravitational potential field using the available DM snapshots (see appendix~\ref{potential}). Subsequently, \ligerGAL employs the merger tree and the gravitational potential to produce a halo catalogue in redshift space by evaluating Eqs.~\eqref{eq:shift_lig}, ~\eqref{eq:deltalna}, and ~\eqref{eq:Mag_lig}, in accordance with the \liger method, i.e., by interpolating the redshift-space position of each particle to the intersection between the halo worldline and the observer’s past light cone.
Lastly, for a given HOD model, we populate each halo with galaxies, correcting for their peculiar velocities by applying the shifts described in Eqs.~\eqref{v_shift_gal} and~\eqref{v_mag_gal} to their positions and magnifications. This procedure, detailed in Sect.~\ref{galaxy_sample}, yields the final galaxy catalogue.
The right side of Fig.  \ref{fig:liger_flowchart} shows the flowchart of this procedure.

Similar to previous applications of the \liger method, we construct series of mock catalogues that include subsets of the full set of relativistic RSDs, which allows us to assess the importance of individual corrections. The catalogues  are  labelled as follows.

The label $\mathcal{R}$ denotes real-space catalogue, where no RSDs are included, $\mathcal{V}$ implies that only the peculiar-velocity terms of the tracers are applied. 
$\mathcal{G}$ indicates that all effects are applied, however, the observer is assumed to be at rest with respect to the cosmic microwave background rest frame, which we will label as Cosmic Rest Frame (CRF), i.e. $\bs{\varv}_o = 0$. 
Finally  $\mathcal{O}$: all effects are applied including the observer's velocity terms, which in this work we assume to have magnitude and direction as measured by the Planck satellite \citep{2020}.

\section{High-Res Mocks}
\label{Methods}
In this section, we discuss our application of \ligerGAL on the HMDPL simulation \citep[][]{2013AN....334..691R, 2012MNRAS.423.3018P, 2016MNRAS.457.4340K}, which we use to generate a halo catalogue in redshift space with the full imprint of linear relativistic RSDs.  
We then describe how we populate this halo catalogue with galaxies, generating a Luminous Red Galaxy mock catalogue that is consistent with our relativistic treatment.

\subsection{HMDPL simulation}
\label{HugeMDPL}

Simulation boxes large enough to encompass the full footprint of Stage-IV galaxy surveys often lack the resolution necessary to support the construction of semi-analytic galaxy catalogues.
One possible solution to circumvent that problem is to replicate smaller boxes to cover larger volumes. However, replications can introduce artificial periodicity, leading to spurious features in the clustering signal - particularly at the large scales relevant for probing relativistic effects \citep{2025arXiv250712116E}.
In this work, we instead follow an approach based on the construction of DM halo lightcone catalogues, which are then populated with galaxies using
the Halo Occupation Distribution  framework \citep[HOD,][]{Berlind_2002}. 
Specifically, we make use of the HMDPL simulation, part of the MultiDark simulation suite \citep[][]{2013AN....334..691R, 2012MNRAS.423.3018P}. HMDPL is a DM–only simulation with a box size of $4\,\mathrm{Gpc}/h$, comprising $4096^3$ particles with a mass resolution of $m = 7.9 \times 10^{10} \, \mathrm{M}_\odot/h$. The simulation adopts a Planck 2014 cosmology \citep[][]{2014} with $(\Omega_{\rm m}, \Omega_{\rm b}, \Omega_\Lambda, \sigma_8 , n_s , h) = (0.307, 0.048, 0.693, 0.829,0.96, 0.678)$. 

From the HMDPL simulation, we import the two available DM particle snapshots at the redshifts $z=0.49$ and $z=0.00$ as well as the ROCKSTAR halo catalogues \citep[][]{Behroozi_2012} from the CosmoSim \citep[][]{2013AN....334..691R} database. The halo catalogues are available in a series of snapshots, roughly uniformly distributed in the scale factor $a$ in the range $0.1\leq a \leq 1 $ (corresponding to $8.58 \geq z \geq0$).

We construct halo trajectories from the merger tree by tracking them across snapshots accounting for mergers and interruptions. Once each trajectory is identified, we determine its intersection with the observer's light cone and use a cubic spline to interpolate the computed shifts and magnifications at the intersection point. We use the virial mass definition of \citet{1998ApJ...495...80B} as a proxy for the halo mass.
We consider all halos with masses that exceed the threshold mass $M_\mathrm{cut}=10^{12}\,\mathrm{M}_\odot/h$ at any redshift\footnote{This requires some care, as a halo near the threshold may initially have a lower mass and later grow to exceed $M_\mathrm{cut}$. To account for such cases, we first apply a lower threshold of $ 5 \times 10^{11}\, \mathrm{M}_\odot/h$, identify the trajectories of all haloes, and then impose the final mass cut of $10^{12}\, \mathrm{M}_\odot/h$.} (see Appendix \ref{lightcone} for further details). We denote the resulting fullsky halo lightcone by \dmcatfs{}. 

The other ingredient necessary for \ligerGAL is the gravitational potential field. Since the HMDPL simulation provides only two DM snapshots, we compute the gravitational potential at these snapshots using spectral methods and employ linear theory to evolve it temporally to any redshift required by the code (see appendix \ref{potential} for  more details and validation of this approach). 
After constructing the trajectories of all halos and calculating the gravitational potential field, we employ \ligerGAL to generate four halo catalogues that incorporate relativistic RSDs at varying levels (see Sect.~\ref{liger4gal}). In addition to the galaxy catalogues derived from these halo catalogues (see Sect.~\ref{galaxy_sample}), we also construct a mass-selected halo catalogue (see Appendix~\ref{lightcone} for details), used in the validation exercises (see Sect.~\ref{validation}). 

\subsection{Galaxy assignment}
\label{galaxy_sample}

In this section, we describe how these halos are then populated with galaxies. As relativistic effects become relevant on scales larger than the size of DM haloes, a Newtonian treatment of this process is sufficient. The key requirement is to maintain consistency with the relativistic RSDs treatment of \ligerGAL, which primarily requires accounting for the additional distortions arising from the fact that populated galaxies do not necessarily share the exact velocities or line-of-sight positions of their parent halos.

We  populate the halo catalogues with luminous red galaxies (LRGs), as their observed distribution in the DESI spectroscopic survey mostly lies within the redshift range covered by our halo catalogues \citep[][]{Adame_2025}.  
For the LRG sample we adopt the  HOD model used by the DESI collaboration \citep[][]{Yuan:2023ezi}.
This HOD model \citep[originally introduced in][]{Zheng_2005} relates the mean occupation of central galaxies to the mass of a given halo,  $M$, via 
\begin{equation}
\label{HOD_Zheng_cen}
    \bar{n}^{\mathrm{LRG}}_\mathrm{cen}(M) = f_\mathrm{ic}\frac{1}{2}\mathrm{erfc}\left[\frac{\mathrm{log}_{10}(M_\mathrm{cut}/M)}{\sqrt{2}\sigma_\mathrm{logM}}\right]\,,
\end{equation}
while the  satellite occupation is given by
\begin{equation}
    \bar{n}^{\mathrm{LRG}}_\mathrm{sat}(M) = \left[\frac{M-M_0}{M_1}\right]^{\alpha}\bar{n}^{\mathrm{LRG}}_\mathrm{cen}(M)\,.
\end{equation}
Here, the HOD model is characterized by six parameters, ${f_\mathrm{ic}, M_\mathrm{cut}, \sigma_{\mathrm{logM}}, \alpha, M_1, M_0}$, whose values are taken from the vanilla model fit listed in Table~3 of \citet{yuan2023desionepercentsurveyexploring}.  These values are extracted from the DESI 1\% data at the redshift bins $z\in[0.4,0.6]$ and $z\in[0.6,0.8]$. We, however,  interpolate their values for each redshift when populating the halo lightcone\footnote{ We apply the parameter $f_\mathrm{ic}$, which accounts for potential incompleteness in the dataset by effectively reducing the average occupation number of each halo, slightly differently in our pipeline. Rather than modifying the average occupation directly, we implement incompleteness by randomly subsampling the galaxy sample after generation, retaining only a fraction $f_\mathrm{ic}$ of the galaxies}, following the approach described in \citet{hadzhiyska2023syntheticlightconecatalogues}. We stress that for the procedure of populating haloes with galaxies we rely on the cosmological redshifts of the haloes, not the observed ones, as in this context $z$ is interpreted as a time coordinate of the Universe evolution, and not as an observable.

After assigning a central galaxy and a number of satellite galaxies to each halo, we generate their phase-space (precisely, coordinate and velocity) positions as follows. Each central galaxy is assigned the same phase-space position as its host halo. For the satellites, we assume an isotropic spatial distribution within the halo, following a Navarro-Frenk-White \citep[NFW,][]{1997ApJ...490..493N} radial profile  without bias. 
The mass–concentration relation is adopted from \cite{2021MNRAS.506.4210I},  via the COLOSSUS code \citep{2018ApJS..239...35D}. For simplicity, we further assume that the satellites are in virial equilibrium within the host halo potential and follow isotropic orbits. 
Under these assumptions, the satellite galaxies'
radial velocity distribution can be characterized by its second moment, which is obtained by solving the Jeans equation \citep[see Eq. 24 in ][]{More_2009}. Allowing us to assign positions $\bs{X}_{\rm g}$
and velocities $\bs{V}_{\rm g}$ to each of the satellite galaxies relative to their respective host haloes. Although the choice of satellite occupation can significantly affect the clustering signal on scales where the one-halo term is important \citep[impacting both the overall clustering and nonlinear RSDs, see e.g.,][]{COORAY_2002}, it has negligible effect on the large-scale regime where the two-halo term dominates, which is the primary focus of this work.
We note, however, that the consistency of the \ligerGAL pipeline is independent of this stage in the catalogue production. 
In principle, more sophisticated methods for assigning phase-space positions can be easily incorporated and implemented. In fact, the numerical implementation of the HOD on the \ligerGAL catalogues relies heavily on the \texttt{HaloTools} library \citep{Hearin_2017}, which allows for the straightforward implementation of complex occupation models.

\begin{figure}
    \centering
    \includegraphics[width=1.\linewidth]{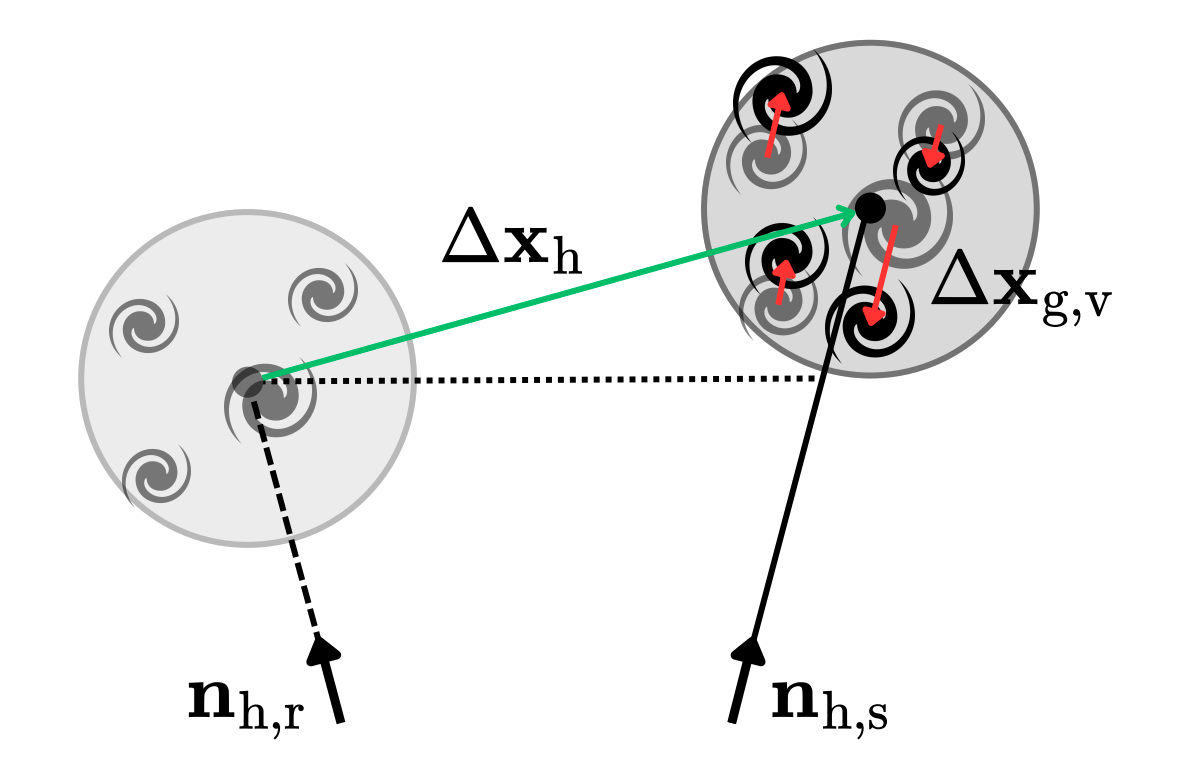}
    \caption{\justifying
    Schematic summarising how the haloes (Section \ref{liger4gal}) and then galaxies (Section \ref{galaxy_sample}) are shifted to build the light cones. 
    We first apply the shift $\Delta \bs{x}_{\mathrm{h}}$ (shown with a green arrow, see Eq.~\ref{eq:shift_lig}) to the halo positions, including both local and integrated terms. Then, we perform the peculiar velocity correction $\Delta \bs{x}_\mathrm{g,v}$ (red arrows) described by Eq.(\ref{v_shift_gal}) to each galaxy. 
    In an analogue way, we produce a global magnification term $\mathcal{M}_\mathrm{h,s}$ for each halo (see Eq.~\ref{eq:Mag_lig}), that is applied to each galaxy inhabiting it after correcting for the individual galaxy peculiar velocities with Eq.(\ref{v_mag_gal}).}
    
    \label{fig:Liger_scheme}
\end{figure}

Equipped  with $\bs{X}_{\rm g}$ and $\bs{V}_{\rm g}$, we  correct the observed positions of satellite galaxies using 
\begin{equation}
\begin{split}    
\label{v_shift_gal}
    \bs{x}_{\mathrm{g},\mathrm{s}} =&\bs{x}_{\mathrm{h},\mathrm{s}}+\bs{X}_\mathrm{g}-\frac{(\bs{n}_{\mathrm{h},\mathrm{s}}\cdot\bs{\varv}_\mathrm{h})}{\mc{H}}\,\bs{n}_{\mathrm{h},\mathrm{s}}\\
    &+\frac{\left[\bs{n}_{\mathrm{g},\mathrm{s}}\cdot(\bs{V}_\mathrm{g}+\bs{\varv}_\mathrm{h})\right]}{{\mc{H}}}\,\bs{n}_{\mathrm{g},\mathrm{s}}\,.
    \end{split}
\end{equation}
where the subscripts $g$ \& $h$ denote the galaxies and halos, respectively. In particular, since this correction is purely radial, the observed galaxy angular positions in the sky $\bs{n}_{\mathrm{g,s}}$ can be obtained before the shift is applied, i.e.,
\begin{equation}
    \bs{n}_\mathrm{g,s}=\frac{\bs{x}_{\mathrm{h},\mathrm{s}}+\bs{X}_\mathrm{g}}{|\bs{x}_{\mathrm{h},\mathrm{s}}+\bs{X}_\mathrm{g}|}\,.
\end{equation}
Similarly, for the satellite galaxy magnification, we implement the correction
\begin{equation}
\begin{split}
\label{v_mag_gal}
    \mathcal{M}_{g,\mathrm{s}} =& \mathcal{M}_{\mathrm{h},\mathrm{s}}+2\left(1-\frac{c}{\mathcal{H}\,|\bs{x}_{\mathrm{h},\mathrm{s}}|}\right)\frac{\bs{n}_{\mathrm{h},\mathrm{s}}\cdot\bs{\varv}_\mathrm{h}}{c} \\
   &- 2\left(1-\frac{c}{\mathcal{H}\,|\bs{x}_{\mathrm{g},\mathrm{s}}|}\right)\frac{\bs{n}_{\mathrm{g},\mathrm{s}}\cdot(\bs{V}_\mathrm{g}+\bs{\varv}_\mathrm{h})}{c}\,.
    \end{split}
\end{equation}
In deriving Eqs.~\eqref{v_shift_gal} and ~\eqref{v_mag_gal}, we assume that the potential-related terms are the same for all members of a given halo. We then account for the velocity terms by removing only the halo peculiar velocity contributions from Eqs.~\eqref{eq:shift_lig} and~\eqref{v_mag_gal}, to then re-add them with the  velocity and line of sight of the member galaxy.
We show a schematic of this procedure in Fig. ~\ref{fig:Liger_scheme}. Since in this work we do not probe scales smaller than $\approx 40\,\mathrm{Mpc}/h$, and the maximum virial radius of haloes appearing in the simulation is $\approx 2.5\, \mathrm{Mpc}/h$, this is an appropriate approximation. 
We note that studies targeting local relativistic effects, such as the gravitational redshift \citep[see, e.g.,][]{Euclid:2024azy}, require more sophisticated modelling of the local gravitational potential terms associated with the galaxies, explicitly accounting for the internal halo density profile. However, such corrections are not essential for the present work, which focuses on large-scale clustering statistics through power-spectrum multipoles, whereas gravitational redshift was found to give a significant contribution at scales around $10\,\mathrm{Mpc}/h$ \citep[][]{Euclid:2024azy, Alam:2017izi}. We show in appendix \ref{gal_approx} a derivation to further support the approximation.

We incorporate magnification bias through a weighting scheme applied to all galaxies \citep[][]{hadzhiyska2023syntheticlightconecatalogues}. \footnote{Another possible approach is to explicitly sample galaxy luminosities \citep[e.g., through subhalo abundance matching techniques,][]{2004ApJ...609...35K}, compute the magnified flux, and then apply a flux cut. For a simple flux-limited selection, both approaches yield the same effect on clustering statistics. The more detailed approach becomes relevant only for complex selection functions.} 
Specifically, for a given population with magnification bias $\mathcal{Q}$, each galaxy is assigned a weight
\begin{equation}
\label{wmag}
w_{\rm g} = \left[1 + \big(\mathcal{M}_{\rm g,\mathrm{obs}} - 1\big)\,\mathcal{Q}\right] ,
\end{equation}
when computing the $N$-point statistics of the sample. We note that this weighting approach differs from the \ligerDM approach used in Eq.~\eqref{eq:delta_g_bc}. The latter relates the magnification field \citep[computed from DM particles using a mass-assignment scheme, see appendix A2 in  ][]{Elkhashab_2021} to the observed overdensity \textit{field}, whereas in our case the weights are applied directly to individual galaxies. 

As for the magnification bias function $\mc{Q}$, we adopt the values for the LRG sample reported in \citet{Zhou_2023}\footnote{\citet{Zhou_2023} define the magnification bias as $s = \mathrm{d}\,\log_{10} N_{\mathrm{g}} / \mathrm{d}m$, where $m$ is the apparent magnitude. That parametrization can be related to our definition via $\mathcal{Q} = 5/2\,s$.} for the two redshift bins, following the same procedure as for the HOD parameters.

\begin{figure}

\centering
    \includegraphics[width=\linewidth]{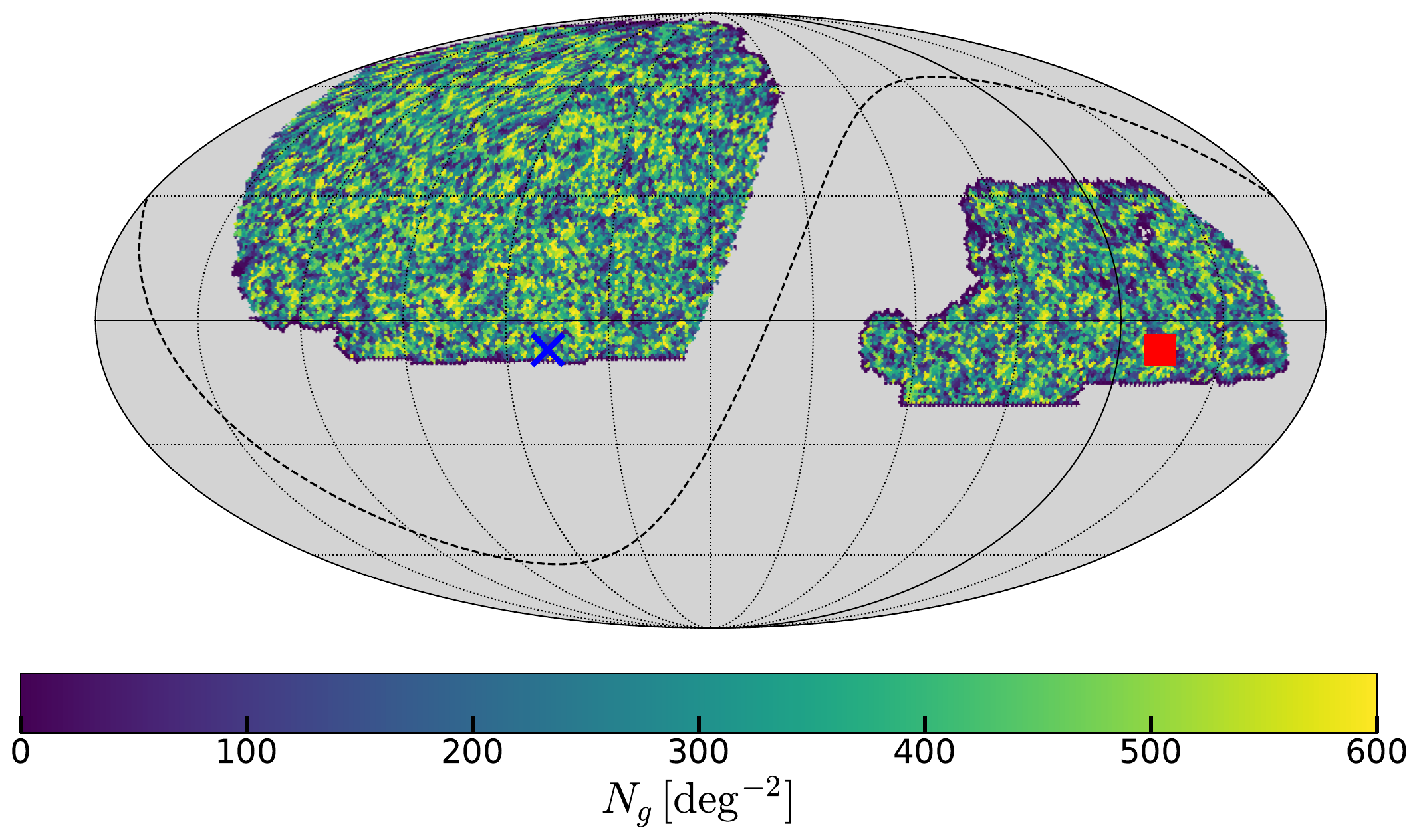}
\caption{\justifying Mollweide projection in ecliptic coordinates of the DESI Year-5 like sample we generated, used to construct the \galcatdesi and \bcgalcatdesi catalogues. The blue cross marks the direction of the observer’s velocity adopted in this work \citep{2020}, while the opposite direction is indicated by the red square. The galactic plane is shown as a dashed black line.}
\label{fig:DESI_mask}
\end{figure}
At the end of this procedure, we obtain four full-sky LRG-like galaxy catalogues in the redshift bin $z\in[0.4,0.8]$ which include varying levels of relativistic RSDs while preserving the nonlinear clustering accuracy of the HMDPL simulation. We use the LRG galaxy catalogues to produce DESI-like ones by applying the angular mask shown in Fig.~\ref{fig:DESI_mask} as well as accounting for the  incompleteness parameter $f_{\rm ic}$ which is set to one in the full-sky case. For clarity, we adopt the following nomenclature to differentiate the catalogues: TYPE–MASK–RSDs. The first entry specifies the type of tracer (H for the halo catalogue, and LRG for galaxies), the second denotes the geometrical footprint (FS for full-sky and DESI for the DESI footprint), and the third indicates the type of RSDs applied, with RSDs $\in \{\mc{O}, \mc{V}, \mc{G}, \mc{R}\}$ (see Sect.~\ref{liger4gal}).    

To ensure that our catalogues accurately depict a DESI-like LRG sample, we compare various properties of the \galcatfs to the LRG sample published  in the  DESI 1\% data 
\citep[see Table 3 in][]{yuan2023desionepercentsurveyexploring}.
The comparison is presented in Table~\ref{tab:subset_table}. Both the linear bias parameter $b_1$ (see appendix ~\ref{LRG_bias} for how the parameter is estimated) and the satellite fraction \(f_{\mathrm{sat}}\) are compatible between the two DESI 1\% data and the \galcatdesi catalogues. 
For the average host mass $\log_{10}\bar{M}_{\mathrm{h}}$, our values are slightly larger than those of the DESI 1\% data, in particular in the later bin. This discrepancy is most likely sourced by the mass resolution of the HMDPL simulations that lead to an underestimation in the Halo Mass function (HMF) in the low-mass end (see appendix ~\ref{lightcone}).
As for the average number density, after accounting for the sample incompleteness, we find that  \(\bar{n} \approx 5.4\times 10^{-4}\,h^{3}\,\mathrm{Mpc}^{-3}\) for the \galcatdesi catalogues, which is consistent with typical expectations for this redshift range \citep[see, e.g.,][]{Zhou_2023}.
Alongside the \ligerGAL version of the code, we make available the toolkit that implements the galaxy generation pipeline: starting from a \liger halo catalogue as input, it produces tracer populations, allowing the selection of different HOD models.

\begin{table*}
\centering
\centering
\begin{tabular}{l c c c c c c}

Redshift bin&Model & $\bar{n}_{\mathrm{g}}\,[h^3\,\mathrm{Mpc}^{-3}]$ & $b_1$ & $f_{\mathrm{sat}}$ & $\log_{10}\left(\bar{M}_{\mathrm{h}}/(M_\odot/h)\right)$ \\
\hline

 \multirow{2}{*}{$z\in[0.4,0.6]$}  &DESI 1\% & --- & $1.94^{+0.04}_{-0.04}$ & $0.089^{+0.013}_{-0.010}$ & $13.42^{+0.02}_{-0.02}$\rule{0pt}{2.6ex} \\
 &\galcatfs& $5.5 \times 10^{-4}$ & $1.90$ & $0.100$ & $13.46$ \rule{0pt}{2.6ex}\\
 \noalign{\vskip 4pt}
\\
  \multirow{2}{*}{$z\in[0.6,0.8]$} & DESI 1\% &  --- & $2.11^{+0.03}_{-0.04}$ & $0.104^{+0.013}_{-0.010}$ & $13.26^{+0.02}_{-0.02}$  \rule{0pt}{2.6ex}\\
&\galcatfs  
 & $6.4 \times 10^{-4}$ & $2.01$ & $0.122$ & $13.34$ \rule{0pt}{2.6ex}\\
 \noalign{\vskip 4pt}

\end{tabular}

\caption{
\justifying The DESI 1\% LRG sample parameters \citep[as reported in Table~5 of][]{yuan2023desionepercentsurveyexploring} are compared to the values derived from our mock catalogues (LRG-FS). Specifically, we report the linear bias, $b_1$, satellite fraction  $f_{\mathrm{sat}}$, and the average halo mass per galaxy $\log_{10} \bar{M}_{\mathrm{h}}$ (in units of \({\mathrm{M}}_\odot / h\)). Additionally, for the mock catalogue, the average galaxy number density \(\bar{n}_{\mathrm{g}}\) in each redshift bin is provided.}

\label{tab:subset_table}
\end{table*}

\subsection{Low resolution light cones}
\label{Monofonic}
To compare the \ligerGAL version of the code with the \ligerDM version, we generate a set of light cones following the procedure described in Sect.~\ref{lowres}. For the required DM-only simulations, we utilise the \textsc{MONOFONIC} code \citep{Michaux_2020} to  produce the simulation snapshots using third-order Lagrangian perturbation theory (3LPT).
We run $64$ simulations, adopting the same cosmological parameters as the HMDPL run (see Sect.~\ref{HugeMDPL}), each using $256^3$ particles of mass $m_{\mathrm{dm}} = 3.25 \times 10^{14}\,\mathrm{M}_\odot/h$ within a periodic box of side length \(4000\,\mathrm{Mpc}/h\). For each simulation, snapshots are stored at the same redshifts as in the HMDPL run, to ensure consistency with the HMDPL catalogues.

We then use \ligerDM to generate a mock catalogue of the halo sample (hereafter \bcdmcatfs{})\footnote{We add the label BC (for \texttt{buildcone}) to differentiate between the low-resolution and high-resolution catalogues.}, and of the LRG galaxy sample. We estimate the survey functions of the two tracers from their number counts power spectrum and radial selection function, following the approach described in Appendix \ref{survey_est}.
In Fig.  \ref{fig:survey_functions} we show the survey functions for the \bcdmcatfs{} halo sample (grey dotted line), the \galcatfs  (black continuos line), and \galcatdesi (red dashed line) catalogues.
The halo catalogues do not constitute a flux-selected sample, so the selection function $\bar{n}(z)$ does not depend on any luminosity cut $L_\mathrm{cut}$. As a consequence, the magnification bias is zero for the halo catalogues.

\begin{figure}

\centering
    \includegraphics[width=\linewidth]{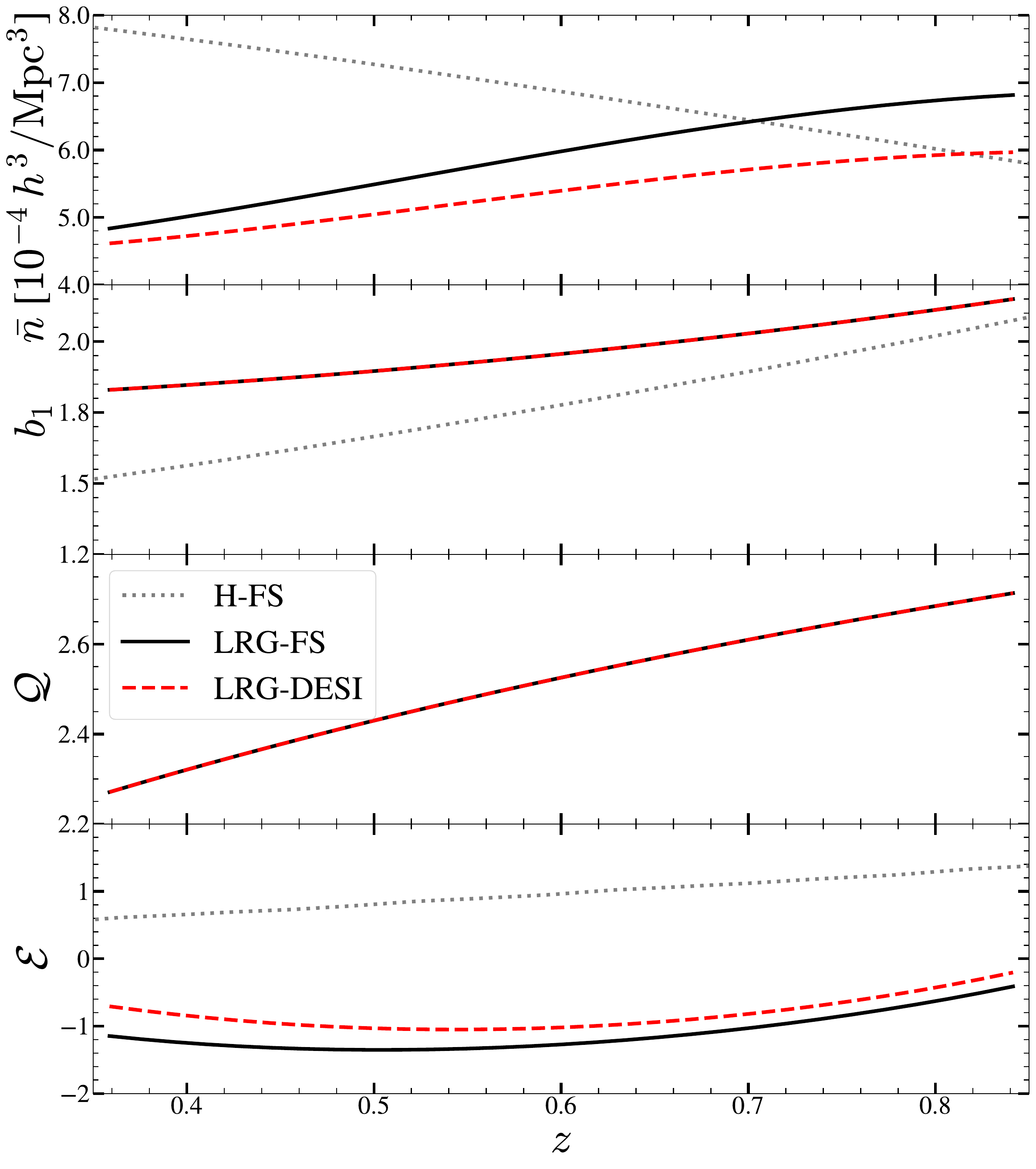}
\caption{\justifying Halo and LRG galaxies survey functions. With a grey dotted line we show the radial density and the linear and evolution biases of the full sky halo sample H, while with a black continuous line and a red-dashed line we show the same quantities along with the magnification bias, for the LRG-like samples in the full-sky (FS) and DESI setting respectively. The radial densities of the two LRG samples differ due to the application of the incompleteness factor of Eq.~\eqref{HOD_Zheng_cen} only to the latter. As mentioned in Sect. \ref{validation}, the halo magnification bias is null, thus not shown.}
\label{fig:survey_functions}
\end{figure}

\section{Results}
\label{Results}
In this section, we outline the main results of this work. Firstly, we quantify the improvements of our \ligerGAL implementation through a comparison with the \ligerDM method and with theoretical predictions. 
Secondly, by performing a standard $f_\mathrm{nl}$ inference from the power spectrum of our galaxy catalogues, we quantify the bias in the measurements of local primordial non-Gaussianity that arises from neglecting relativistic RSDs and - in particular - our peculiar velocity in the theoretical modelling of the galaxy power spectrum.
Lastly, we perform a full-shape analysis of the power spectrum to assess the extent of the bias  sourced by relativistic RSDs in the inference of cosmological parameters in a $\Lambda \mathrm{CDM}$ universe. 

\subsection{Validation}
\label{validation}

\begin{figure*}
    \centering
    \includegraphics[width=1.\linewidth]{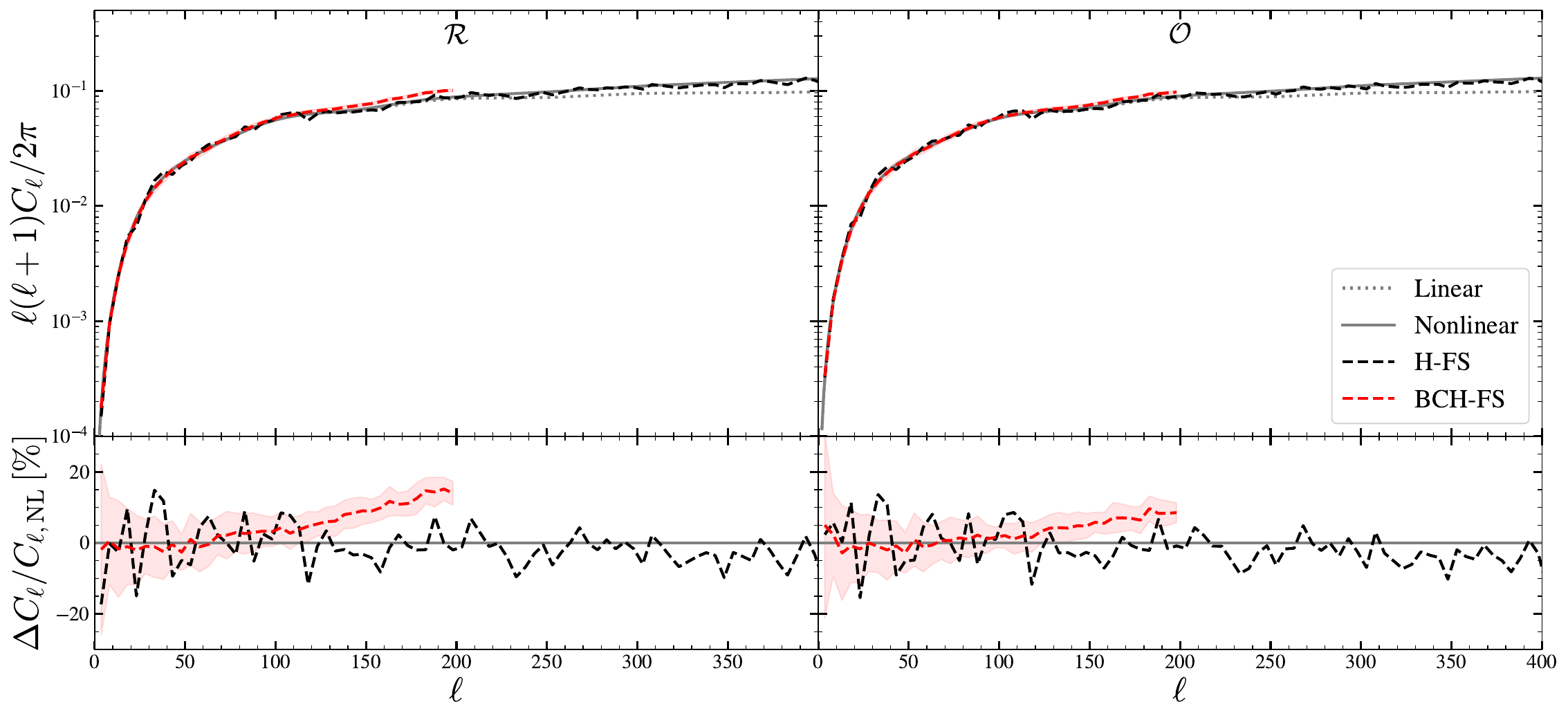}
    \caption{\justifying \textit{Upper :} We show the halo angular power spectra of the \dmcatfs{} catalogues (black dashed lines) and of the {low-resolution} \bcdmcatfs{} catalogues, where mean of the mocks is shown with a red dashed line and the R.M.S. with a  pink-shaded region). Spectra are shown for the $z \in [0.5,0.6]$ bin, with $\mathcal{R}$ and $\mathcal{O}$ mocks in the left and right panels. CAMB linear (grey dotted lines) and halofit (grey dash-dotted) predictions are overplotted.
\textit{Lower :} Residuals relative to the CAMB non-linear prediction.}
    
    \label{fig:cl_comparison}
\end{figure*}
\subsubsection{Angular clustering}
\label{angular_clustering}
Following the approach adopted  in \citet{Elkhashab_2021}, we use the angular power spectrum, $C_\ell$, to validate the clustering properties of the \ligerGAL method against the \ligerDM approach as well as against theoretical predictions, which we calculate using the CAMB \citep{2011ascl.soft02026L} code. For a validation of the \ligerDM method, we refer to previous works \citep[][]{Borzyszkowski:2017ayl,Elkhashab_2021,2025}.
We measure the $C_\ell$ for both the \dmcatfs{} and \bcdmcatfs{} halo samples in the redshift bin $z\in[0.5,0.6]$ , using the pseudo-$C_\ell$ (PCL) estimator \citep[][see also~\ref{cl_estimator} for a brief summary]{1973ApJ...185..413P}.
We consider a full-sky survey, generating HEALPIX maps with $12\times1024^2$ pixels, which allows us to access the $C_\ell$ far beyond the scales of interest.
For the CAMB prediction, we include  linear biasing as well as all linear relativistic RSDs (the latter only for comparison with $\mc{O}$ mocks) through the survey functions of \bcdmcatfs{} shown in Fig.~\ref{fig:survey_functions}.

We show the  $C_\ell$ comparison  for the  $\mathcal{R}$ (left panel) and $\mathcal{O}$ (right panel) mocks in Fig.  \ref{fig:cl_comparison}.
 The top panels show the $C_\ell$ predicted by the CAMB linear model (grey dotted line) and its non-linear \citep[][]{Mead_2021} counterpart. Theoretical lines are compared with the spectra estimated from the \dmcatfs{} (black dashed line) and \bcdmcatfs{} catalogues. For the latter, we display the average spectrum with its R.M.S. scatter (red dashed line and pink shaded region). The bottom panels show the residuals with respect to the non-linear theoretical prediction.
In both cases, the large-scale signal is consistent between the two \liger implementations, showing residuals comparable to the scatter of the \bcdmcatfs{} catalogues. 
At smaller scales, the \ligerDM approach overestimates the clustering signal due to the limited resolution of the linear biasing procedure on which it relies, whereas the \ligerGAL implementation continues to provide an accurate description.
As we reach $\ell=300$, the non-linear model also ceases to reproduce the halo power spectrum, as expected.

\subsubsection{The FOTO signal}
\label{foto_sec}

The FOTO signal is an oscillatory imprint in the power spectrum sourced by the $\varv_{\rm o}$ term in Eq.~\eqref{eq:Deltag}. Its contribution to the monopole of the power spectrum, accounting for a full-sky geometry, is given by \citep{Elkhashab_2021,Elkhashab_2025}
\begin{equation}
\label{foto}
    P_{0,\mathrm{FOTO}}(k) = \frac{16\pi^2}{3}\frac{\varv_o^2}{H_0^2}\frac{I_1^2(k)}{4\pi\int_{r_1}^{r_2}r^2\bar{n}_\mathrm{g}^2\,\dif r}\,,
\end{equation}
where 
\begin{equation}
    I_1(k) = \int_{r_1}^{r_2}\frac{r\,\bar{n}_\mathrm{g}\,\alpha_0}{a\,H/H_0}\,j_1(kr)\,\dif r\,,
\end{equation}
with the integration limits corresponding to the boundaries of the redshift shell, and
\begin{equation}
\label{alpha_0}
    \alpha_0(z)=2(1-\mathcal{Q})+\left[3-\mathcal{E}-\frac{\mathrm{d}\,\mathrm{ln}\,H}{\mathrm{d}\,\mathrm{ln\,(1+z)}}\right]\frac{r\,H}{c(1+z)}\,.
\end{equation} 

\begin{figure}
    \centering
    \includegraphics[width=1.\linewidth]{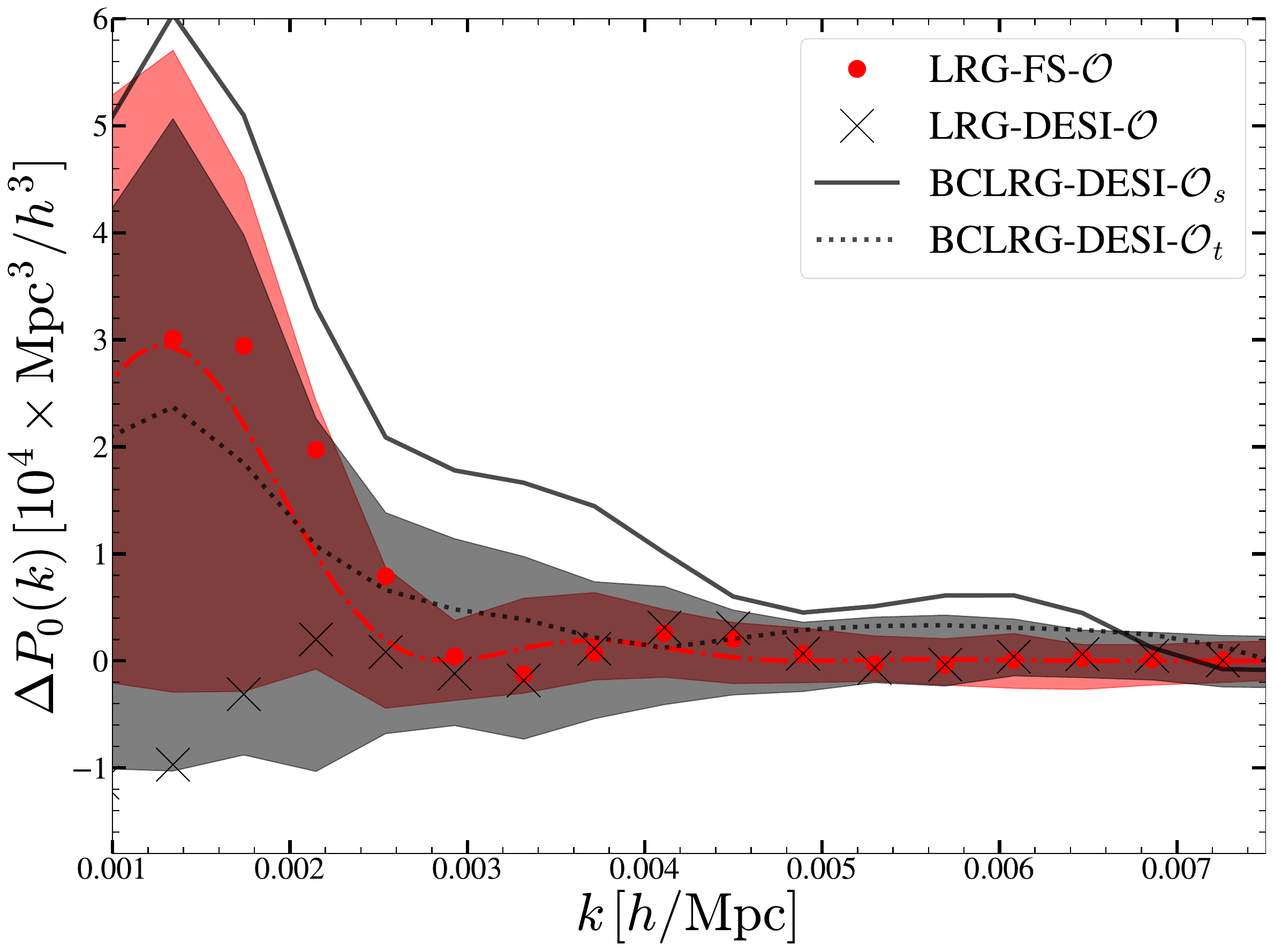}
    \caption{\justifying We plot the FOTO signal for the power spectrum monopole of different LRG samples. The full-sky \galcatfs measurement is shown as red dots, while the $\sigma_{68}$ region derived from the {low-resolution} \bcgalcatfs catalogues is shown as  the pink-shaded area. The corresponding masked-sky case \galcatdesi and \bcgalcatdesi are represented respectively by black crosses and grey-shaded regions. Additionally, a typical and a strong masked-sky FOTO signals  are shown as dotted and solid lines, respectively. Finally, the theoretical prediction for the full-sky case is plotted as a dash-dotted line.}
    
    \label{fig:FOTO_comp}
\end{figure}

We present the FOTO signal for our galaxy catalogues in Fig.~\ref{fig:FOTO_comp}, obtained by taking the difference between the $P_0(k)$ measurements from the $\mathcal{O}$ and $\mathcal{G}$ catalogues, i.e $\Delta P_0 \approx P_{0,\rm FOTO}$. The spectra are computed by embedding the catalogues in a box with side length of $L_{\rm FFT}=16\,000\,\mathrm{Mpc}/h$, ensuring that the large-scale behaviour is adequately captured. We estimate the spectra using the \ttt{pypower} code \citep{adame2024desi} that implements the standard Feldman-Peacock-Kaiser \citep[hereafter, FKP,][see appendix \ref{estimators} for more details]{Feldman_1994,yamamoto_cosmological_1999}. We show the signal of the \galcatfs catalogues as red dots, with the pink-shaded region is bordered by the 16th and 84th percentiles (hereafter, $\sigma_{68}$ region ) of the FOTO signal computed from the \bcgalcatfs catalogues. For the masked-sky case, the spectrum of the \galcatdesi catalogue is shown as black crosses, while the corresponding $\sigma_{68}$ region derived from the \bcgalcatdesi mocks is represented by the grey-shaded region. We also select two realizations from the \bcgalcatfs mocks that exhibit a ``typical" and a strong FOTO signal, illustrated with dotted and solid lines, respectively. For comparison, we also include the theoretical prediction (see Eq.~\ref{foto}) for the full-sky case. It is clear from the figure that the theoretical prediction is consistent with the numerical signal from both catalogues.

The signal has interesting features in the masked sky case. Firstly, the survey mask modifies the oscillatory pattern by shifting both the amplitudes and the peak positions as it mixes between the $k$-modes. Secondly, a higher fraction of the realizations produces a negative FOTO signal w.r.t. the full-sky case, because the survey mask also induces mixing between the multipoles of the power spectrum \citep[e.g.,][]{2017MNRAS.464.3121W}, and the higher multipoles of the FOTO signal, which are not positive definite \citep{Elkhashab_2025}. Additionally, the incompleteness factor $f_\mathrm{ic}$ applied only to masked mocks introduces a different redshift dependence, further modifying the signal.

The comparison clearly indicates a stronger FOTO effect in full-sky mocks, where the mean signal is larger and the uncertainties are smaller, due to the larger volume coverage.
However, the signal of individual realizations in the masked-case varies strongly.
Consequently, for the forthcoming analysis, we adopt the \galcatfs signal as a “typical” scenario, since it is an accurate description of the average signal.
For the masked-sky case, we consider three spectra: the first is computed from  the  \galcatdesiWr{O} catalogue, which exhibits a very weak FOTO signal. In addition, we construct two distinct power spectrum measurements by imprinting the FOTO signature from a ``typical" \bcgalcatdesi realization (i.e. within the $1\sigma$ range of the possible realizations in the scales of interest) and a strong \bcgalcatdesi realization onto the $\mathcal{G}$ catalogue of the \galcatdesi mocks. The power spectrum multipoles after this imprint are then
\begin{equation}
    P^{\mathcal{O}_i}_\ell(k)=P^\mathcal{G}_\ell(k)+\Delta P_\ell^i(k)\,,
\end{equation}
where $P^\mathcal{G}_\ell$ are the power spectrum multipoles of the \galcatdesiWr{G} catalogue, and $\Delta P^\mathrm{i}_\ell$ is the FOTO signals obtained from one of the two \bcgalcatdesi realizations considered.
We denote these two signals as \galcatdesiWr{\mc{O}_t} and \galcatdesiWr{\mc{O}_s}, respectively. In Table \ref{realizations}, we show the different types of  mocks that are used in the analysis of the LRG-DESI sample as well as the relativistic RSDs included in these catalogues (see also Sect.~\ref{liger4gal}).

\begingroup
\renewcommand{\arraystretch}{1.2}
\begin{table}
\centering
\caption{\justifying Different labels for the mocks used in this work. While $\mathcal{R}$, $\mathcal{V}$, $\mathcal{G}$ and $\mathcal{O}$ refer to the different levels of GR corrections that we implement in each sample, $\mathcal{O}_t$ and $\mathcal{O}_s$ indicate two specific realizations of the full treatment $\mathcal{O}$ that we have chosen to analyse more in depth.}
\begin{tabular}{lr}
 Mock & Description\\
 \hline
 $\mathcal{R}$&Real space\\
 $\mathcal{V}$&Peculiar velocities only\\
 $\mathcal{G}$&Full GR in the CRF\\
 $\mathcal{O}$&Full GR with $\bs{\varv}_0$ from \citet{2020}\\
 \\
 $\mathcal{O}_t$&Selected typical realization of $\mathcal{O}$\\
 $\mathcal{O}_s$&Selected strong realization of $\mathcal{O}$\\
 
 \end{tabular}
\label{realizations}
\end{table}
\endgroup

\subsection{ Impact  of Relativistic RSDs on cosmological parameters}
We now turn to the central aim of this work: evaluating the impact of relativistic RSDs on cosmological parameter measurements using the galaxy catalogues we generated. As these effects affect the large-scale galaxy clustering signal, neglecting them in the theoretical modelling of the observables can bias measurements of cosmological parameters and lead to false detection of certain signatures.
First, we show the impact of relativistic effects for a comoving observer, then we focus, in particular,  on the FOTO signal and the bias it induces in estimates of Local PNG, as well as its influence on parameters inferred from the full-shape fit of the power spectrum, when not properly accounted for in the theoretical modelling of the galaxy power spectrum.

\subsubsection{Impact of other relativistic contributions}

\begin{figure}
        \centering
        \includegraphics[width=\linewidth]{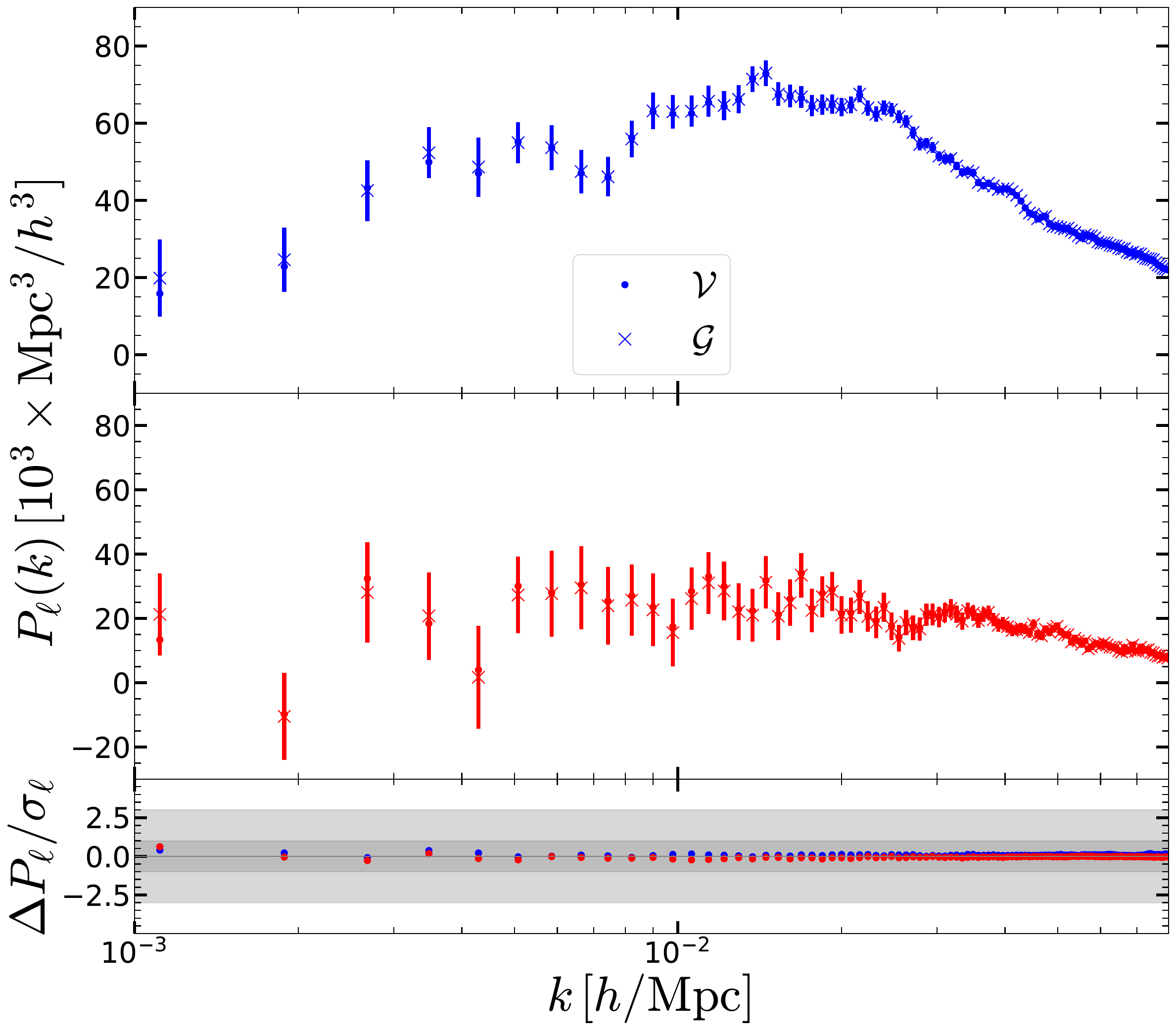}
\caption{\justifying \textit{Upper :} The power spectrum monopole (upper panel) and quadrupole (middle panel) measurements are shown for the full-sky \galcatfs catalogues. The spectra of both the $\mc{G}$ (crosses) and $\mc{V}$ (barred points) are shown. The error bars are identical for  all realizations with the same geometry and  are displayed only for the $\mc{G}$ mocks.
\textit{Lower :} Residuals of the measurements in the $\mc{G}$ and $\mc{V}$ catalogues in units of standard deviation.}
\label{fig:g_test}
\end{figure}

We first examine the impact of relativistic effects on the observed power spectrum multipoles for an observer comoving with the LSS frame ($\varv_{o} = 0$). To this end, we compute the power spectrum multipoles of both the $\mc{G}$
 and $\mc{V}$ mocks from the \galcatfs mocks, shown in Fig.~\ref{fig:g_test}, thereby isolating the contribution of all relativistic RSD terms beyond the peculiar velocities of sources. The error bars represent the square root of the diagonal elements of the covariance matrix, estimated using the \texttt{THECOV} code \citep[][see Appendix~\ref{estimators} for details]{Alves2024prep}. As shown in the figure, the relative discrepancies remain below $1\sigma$ at all scales probed, as expected given the relatively low redshift range of the sample, where the dominant relativistic contribution from gravitational lensing is known to peak at higher redshifts \citep{PhysRevD.84.043516}.

\subsubsection{Local PNG}

In this section, we measure the bias in measurements of local PNG from the observed power spectrum due to relativistic RSDs.
We characterize the deviation from Gaussianity of a random field $\phi_\mathrm{NG}$ by the following parametrization \citep[see e.g.][]{Komatsu_2001}:
\begin{equation}
    \phi_\mathrm{NG}=\phi+f_\mathrm{nl}(\phi^2-\langle\phi\rangle^2)\,,
\end{equation}
where $\phi$ is a Gaussian random field and $f_\mathrm{nl}$ denotes the amplitude of the deviation from a Gaussian distribution.
The local PNG contribution at the initial conditions modifies the observed power spectrum through its impact on the linear bias of tracers of DM \citep[see e.g.][]{Matarrese_2008,Slosar_2008}.  
Our approach to measuring the induced bias in $f_{\rm nl}$ is as follows.  
We extract $f_{\rm nl}$ from the \galcatfs and \galcatdesi catalogues w/o relativistic RSDs (from the $\mc{O}$ and $\mc{V}$ catalogues), using standard modelling that accounts only for velocity distortions. 

Since the base HMDPL simulations (see Sec.~\ref{HugeMDPL}) are generated with Gaussian initial conditions, i.e., $f_\mathrm{nl}=0$, any significant detection of $f_\mathrm{nl}$ in the $\mc{O}$ catalogues w.r.t. the $\mc{V}$ catalogues is attributed to systematic biases due to neglecting relativistic RSDs in the theoretical model. Comparing to the $\mc{V}$ catalogues, rather than directly to the fiducial value, $f_{\mathrm{nl}} = 0$,  allows us to disentangle relativistic RSDs from other possible large-scale  systematics, such as wide angle effects or the radial integral constraint \citep[see e.g.][for methods to account for this effect]{de_Mattia_2019,adame2024desi}. We thus define the quantity 
\begin{equation}
    \Delta f_\mathrm{nl} = f_\mathrm{nl}^{\mathcal{O}} - f_\mathrm{nl}^{\mathcal{V}},
\end{equation} 
where $f_\mathrm{nl}^{\mathcal{O}/\mathrm{V}}$ are measured from the $\mathcal{O}$ and $\mathcal{V}$ catalogues respectively. 

We adopt the following parametrization for the redshift-space power spectrum previously used in  measurements of $f_\mathrm{nl}$ \citep{Castorina_2019,chaussidon2024constrainingprimordialnongaussianitydesi}:
\begin{equation}
\begin{split}   
P^{\rm Th}(k,\mu)=&\frac{\left\{b_1(z_\mathrm{eff})+\left[b_\Phi(z_\mathrm{eff})/T_{\Phi\rightarrow\delta}(k,z_\mathrm{eff})\right]\,f_\mathrm{nl}+f(z_\mathrm{eff})\,\mu^2\right\}^2}{\left[1+\frac{1}{2}(k\,\mu\,\Sigma_{\rm s})^2\right]^2}\\&\times P_\mathrm{lin}(k,z_\mathrm{eff})+s_{\rm n,0}\,{P}^\mathrm{SN}_{0}\,,
\end{split}
\label{pku_ng}
\end{equation}
where  $P_\mathrm{lin}(k,z_{\rm eff})$ is the linear matter power spectrum evaluated at the effective redshift $z_{\rm eff}$,     $T_{\Phi\rightarrow\delta}$ is the transfer function between the primordial gravitational field $\Phi$ and $\delta_{\rm DM, r}$, and $b_\Phi$ is the non-Gaussian contribution to the bias of the tracer. The velocity RSD contribution is included in the model via the  Kaiser correction, $f\,\mu^2$, for RSDs \citep{1987MNRAS.227....1K}, where $f$ is the growth rate of structures, and $\mu$ is the cosine of the angle between the $\bs{k}$-mode considered and a constant line-of-sight. The parameter $s_\mathrm{n,0}$ captures deviations from Poissonian shot-noise and ${P}^\mathrm{SN}_{0}$ is the shotnoise of the measurement (see appendix ~\ref{estimators}). Finally, a finger-of-god (FoG) damping factor is included to partially account for non-linear RSDs  modulated by the parameter $\Sigma_{\rm s}$.

We parametrize $b_\Phi$  as
\begin{equation}
    b_\Phi=2\delta_c \,[b_1(z)-p]\,,
\end{equation}
where $\delta_c=1.686$ is the critical overdensity in the spherical collapse model \citep[see e.g.][]{Sheth_1999}, and $p$ encodes the merger history. {The appropriate value of $p$ remains a subject of debate \citep[see e.g.][]{Barreira_2020}. As this work adopts an HOD that depends solely on halo-mass, we assume $p=1$, following the approach used for the LRG sample in the DESI DR1 analysis by \citet{chaussidon2024constrainingprimordialnongaussianitydesi}.} Regardless of this choice, we will quantify the biases in $f_\mathrm{nl}$ in units of its uncertainty $\sigma_{f_\mathrm{nl}}$. Since both these quantities scale with $1/b_\Phi$, we therefore expect our results to be independent from this assumption.

We estimate $z_\mathrm{eff}$ using
\begin{equation}
\label{z_eff}
    z_\mathrm{eff}=\frac{\sum_i w^2_i\bar{n}_{{\rm g},i}\,z_i}{\sum_i w^2_i\bar{n}_{{\rm g},i}\,}\,,
\end{equation} where  $z_i$ is the redshift of the $i-\mathrm{th}$ galaxy, $\bar{n}_{{\rm g},i}$ is the sample selection function, and $w_i$ are the FKP weights (see Eq.~\ref{w_fkp}), evaluated at the galaxy positions. We note that, in analyses of 
$f_\mathrm{nl}$, the standard FKP weights are often replaced with optimal weights to enhance the constraining power on 
$f_\mathrm{nl}$
 \citep[see][]{Castorina_2019}. However, we adopt the FKP weights in this work, as the redshift-dependent weighting scheme has been shown to have a negligible impact on the $f_\mathrm{nl}$ constraints for the LRG sample \citep[see Figs. 8 and 9a in][]{chaussidon2024constrainingprimordialnongaussianitydesi}.

We account for the different geometries of the \galcatfs and \galcatdesi through the mixing matrix formalism. That approach uses the random catalogue  to calculate a mixing matrix, $\mathcal{W}_{\ell\,{\ell^\prime}}$, which relates the observed power spectrum multipoles to the theoretical model via \citep[][]{euclidcollaboration2024euclidpreparationimpactrelativistic,2014MNRAS.443.1065B,2017MNRAS.464.3121W,Beutler_2021} 
\begin{equation}
\label{conv}
\begin{split}
    P_{\mathrm{obs},\ell}(k)&=\sum_{\ell^{\prime}=0,2,4}\int_0^{\infty}{k^\prime}^2\,\mathcal{W}_{\ell\,{\ell^\prime}}(k,k^\prime)\,P^{\rm Th}_{\ell^\prime}(k^\prime)\,\dif k^\prime\\
    &-\frac{P_\ell^W(k)}{P^W_0(0)}\,\sum_{\ell^{\prime}=0,2,4}\int_0^{\infty}{k^\prime}^2\,\mathcal{W}_{\ell,{\ell^\prime}}(0,k^\prime)\,P^{\rm Th}_{\ell^\prime}(k^\prime)\,\,\dif k^{\prime}\,.
\end{split}
\end{equation}
Where the first term accounts for the mixing between the different modes due to the survey geometry and  the second term corrects for the global integral constraint. We compute the mixing matrices from the random catalogues using the \ttt{CONVO} code \footnote{Available at \url{https://gitlab.com/jacoposalvalaggio/convo}.} (Salvalaggio et al. in prep.). Finally, the theoretical model multipoles are obtained by the Legendre expansion of Eq.~\eqref{pku_ng}
\begin{equation}
    P^{\rm Th}_\mathrm{\ell}=\frac{2\ell+1}{2}\int_{-1}^{1}\,P^{\rm Th}(k,\mu)\,\mathcal{L}_\ell(\mu)\,\,\dif \mu\,.
\end{equation}

\begin{figure*}
    \centering
    \begin{subfigure}[t]{0.49\linewidth}
        \centering
        \includegraphics[width=\linewidth]{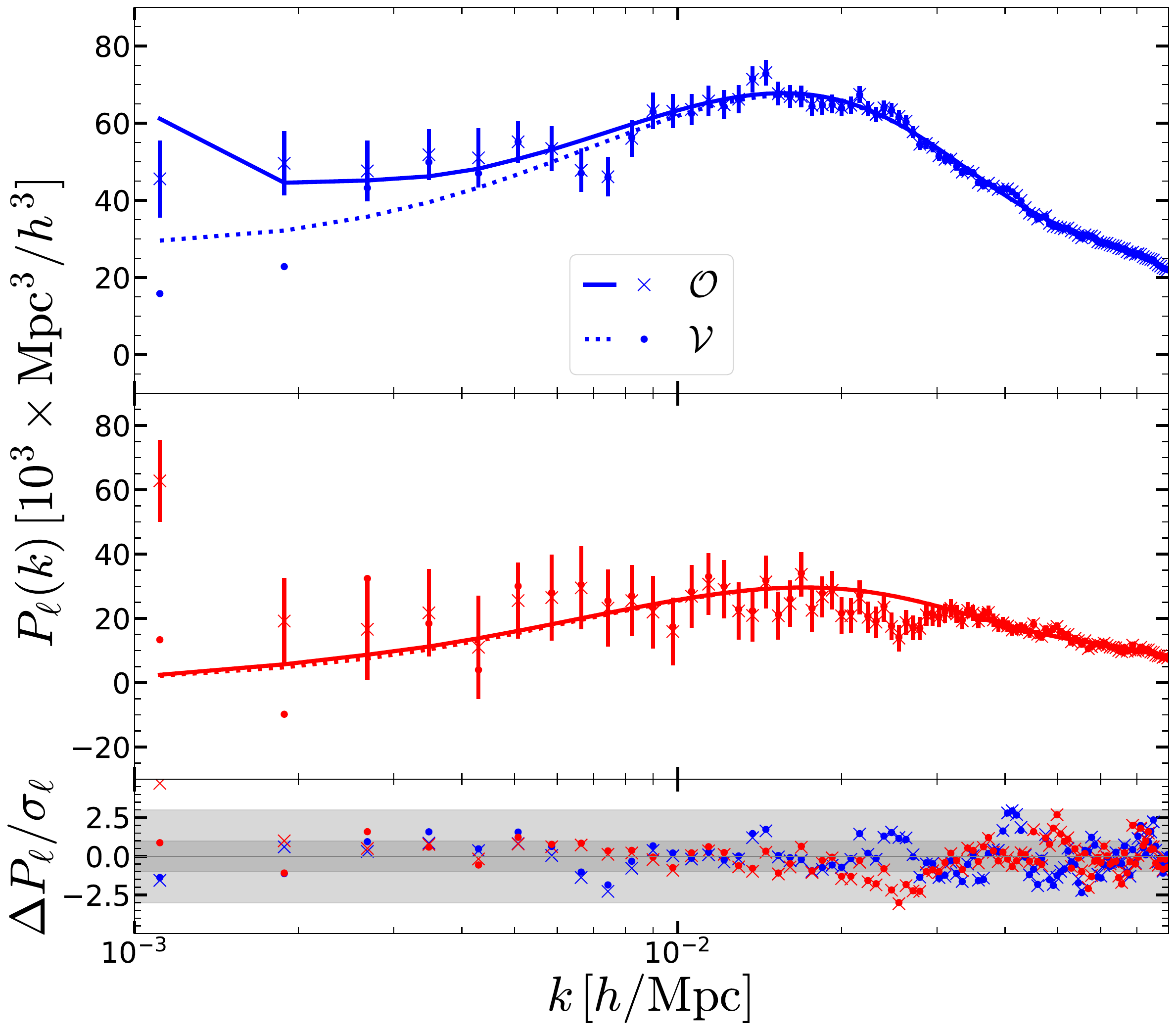}
        \captionsetup{width=0.9\linewidth}
        \caption[width=1.\linewidth]{Full-sky}
        \label{fig:pk_total_kmin}
    \end{subfigure}
    \begin{subfigure}[t]{0.49\linewidth}
        \centering
        \includegraphics[width=\linewidth]{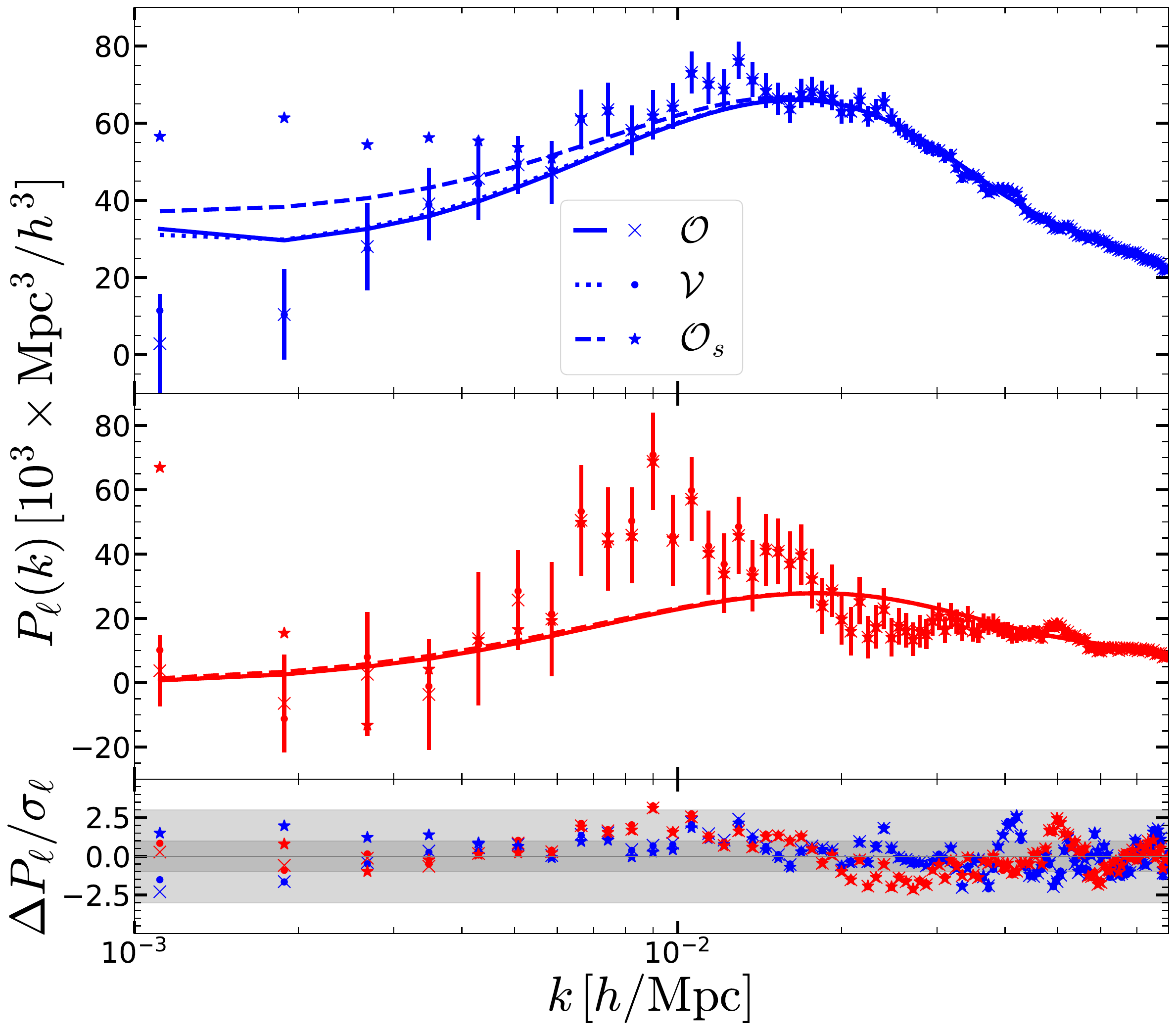}
        \captionsetup{width=0.9\linewidth}
        \caption[width=1.\linewidth]{DESI-like}
        \label{fig:pk_desi_kmin}
    \end{subfigure}
    \caption{\justifying \textit{Upper :} The power spectrum monopole (upper panels) and quadrupole (middle panels) measurements are shown for the full-sky \galcatfs catalogues (left panels) and the masked-sky \galcatdesi catalogues (right panels). The spectra of both the $\mc{O}$ (crosses) and $\mc{V}$ (barred points) are shown as well as the ``strong" \galcatdesiWr{\mc{O}_s} (stars) realization in the partial sky case on the right. The error bars are identical for  all realizations with the same geometry and  are displayed only for the $\mc{O}$ mocks. Solid, dashed, and dotted lines indicate the best-fit models for the $\mathcal{O}$, $\mathcal{V}$, and  $\mc{O}_s$ (right panel only) mocks, respectively. 
    \textit{Lower :} Residuals of the measurements and their corresponding best-fit models in units of standard deviation.}
    \label{fig:pk_comparison}
\end{figure*}
\begin{figure}
    \centering
    \includegraphics[width=1.0\linewidth]{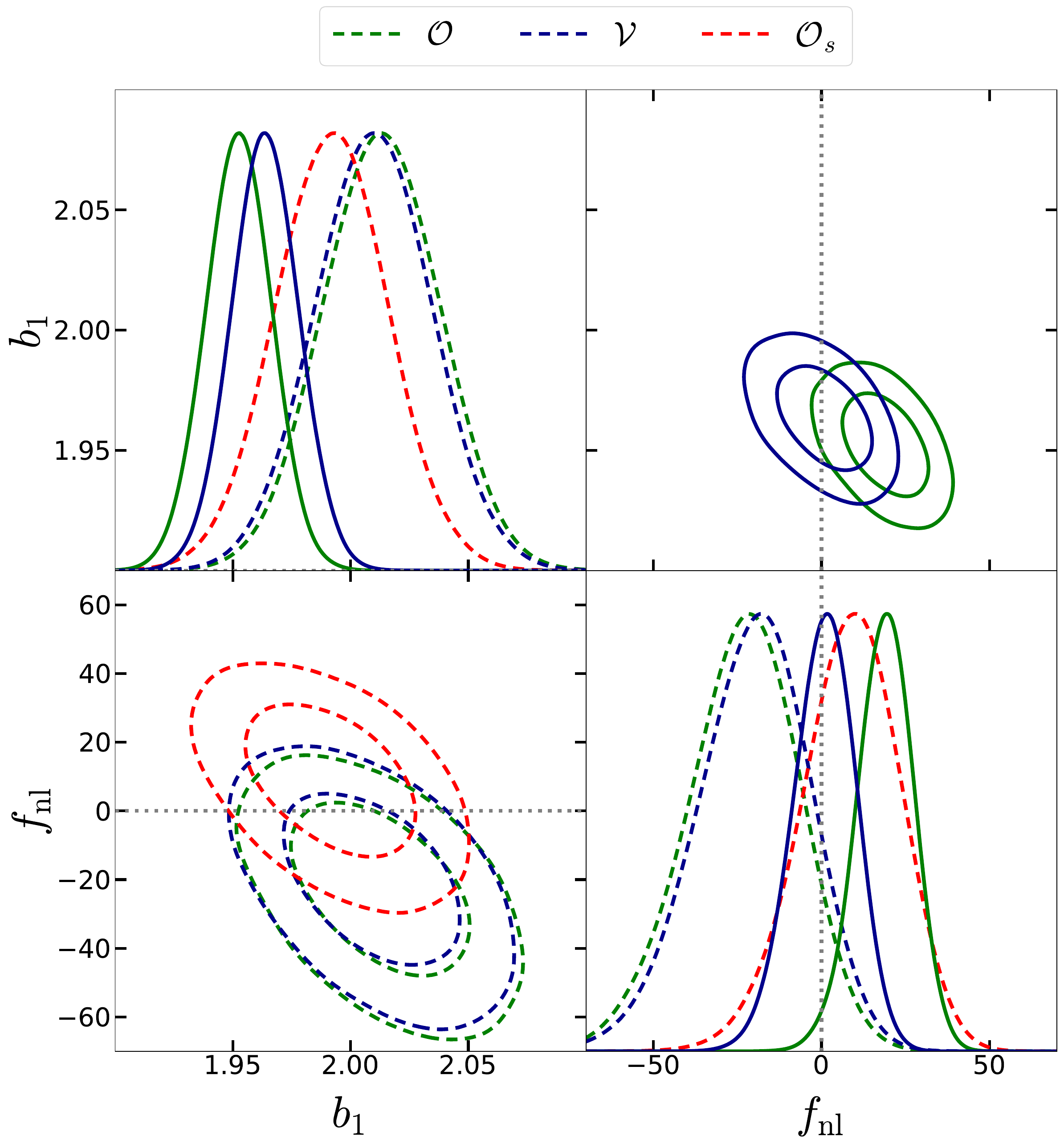}
    \caption{\justifying Constraints on $f_\mathrm{nl}$ and $b_1$ obtained from the full-sky \galcatfs (top right, solid lines) and masked-sky \galcatdesi (bottom left, dashed lines) catalogues, with all relativistic effects (blue) and velocity-only (green) mocks.  
    The fiducial $f_\mathrm{nl}=0$ value is marked by the grey dotted line. Moreover, for the partial sky case, we also show the constraints obtained from the ``strong" \galcatdesiWr{\mc{O}_s} realization in red. }
    \label{fig:fnl_comparison}

\end{figure}

We implement the model of Eq.~\eqref{conv} and fit it to our measurements of the power spectrum monopole and quadrupole, varying four parameters  $\bs{\theta}=[b_1,f_\mathrm{nl},s_{\rm{n},0},\Sigma_p]$. 
For this analysis, we focus on the full redshift bin of our mock catalogue, $z \in [0.4,,0.8]$, as it offers the strongest constraining power.  The power spectra are estimated using an FFT box of size $8{\,}000\,\mathrm{Mpc}/h$, with a $k$-binning of approximately $\Delta k \approx 10^{-3}\,h/\mathrm{Mpc}$. Following \cite{chaussidon2024constrainingprimordialnongaussianitydesi}, we restrict the analysis to scales up to $k_\mathrm{max}=8\times 10^{-2}\,h/\mathrm{Mpc}$, in order to mitigate non-linear effects that may not be adequately captured by the simple FOG model adopted in this work.  For the parameter inference, we employ the importance nested sampling algorithm for Bayesian posterior reconstruction, implemented in the \texttt{Nautilus} package \citep{nautilus}. The resulting parameter chains are then analysed and visualized using the Python package \texttt{getdist} \citep{Lewis:2019xzd}. We assume a Gaussian likelihood and compute the covariance using the \texttt{THECOV} code. Finally,  we adopt wide uniform priors for the model parameters, shown in Table \ref{priors}.  

\begingroup
\renewcommand{\arraystretch}{1.2}
\begin{table}
\centering

\caption{\justifying Priors used in this work for the parameter inference of the primordial non-Gaussianity. The symbol $\mc{U}$ represents a uniform distribution.}
\begin{tabular}{lr}

 Parameter & Prior\\
 \hline
 $b_1$&$\mc{U} (0.25,6.0)$\\
 $f_\mathrm{nl}$&$\mc{U}(-500,+500)$\\
 $\mathrm{s_{\rm n,0}}$&$\mc{U}(-1,2)$\\
 $\Sigma_p\,[\rm{Mpc}\,h^{-1}]$&$\mc{U}(0,20.0)$\\
\end{tabular}
\label{priors}
\end{table}
\endgroup
For the data vectors, we repeat the fits under multiple scenarios. First, we fit the spectra of the the $\mathcal{V}$ and $\mathcal{O}$ realizations of \galcatfs catalogues. Second, we consider the partial-sky \galcatdesi catalogues, analysing the $\mathcal{V}$, $\mathcal{O}$, and $\mathcal{O}_{s}$ realizations. All data vectors employed in the fits are shown in Figure~\ref{fig:pk_comparison}, with the \galcatfs spectra presented in the left panel and the \galcatdesi spectra presented in the right panel. The error bars are derived from the analytical covariance, under the assumption that the covariance is unaffected by the inclusion of relativistic effects. However, the error bars differ between the \galcatfs and \galcatdesi spectra due to the different volumes of the two catalogues. In the \galcatfs case (left panel), we notice that the difference between the $\mathcal{O}$ and $\mathcal{V}$ measurements becomes increasingly evident as we decrease $k$, while for $k> 10^{-2}\,h/\mathrm{Mpc}$ they overlap. In the \galcatdesi on the other hand, the $\mc{O}$ and $\mc{V}$ overlap for most of the k-range, while the $\mc{O}_s$ has a higher signal at small $k$. 
We note an apparent discrepancy between the LRG–DESI measurements and the best-fit model around $k=10^{-2}\,h/\mathrm{Mpc}$. This feature appears consistently in all mock measurements for the masked-sky case but is absent in the full-sky (\galcatfs) case. However, both the masked- and full-sky analyses yield comparable goodness-of-fit values ($\chi^2$), indicating that this deviation is consistent with statistical fluctuations. Furthermore, because the feature persists independently of the level of relativistic effects included, it does not bias our constraints on $f_{\rm nl}$. 

Figure \ref{fig:fnl_comparison} presents the estimated posterior distributions in the $f_{\rm nl} - b_1$ plane. In the  \galcatfs case (upper-right panel), the power spectrum multipoles constrain $f_{\rm nl}$ with a half-width of the 68\% credibility interval of $\sigma_{f_{\rm nl}} \approx 9$. Including relativistic effects in the mocks (comparing the solid blue and solid green curves) induces a bias of $\Delta f_\mathrm{nl} \approx 18 =2\sigma_{f_{\rm nl}}$ in the marginalized posterior means.
For the partial-sky \galcatdesi case (lower-left panel), the $f_{\rm nl}$ parameter is constrained with $\sigma_{f_{\rm nl}}\approx 16$. The weaker constraint results from the reduced sky coverage.
We test two realizations of the FOTO signal: the $\mc{O}$ case (green dashed line), corresponding to a weaker FOTO signal (see blue crosses in Fig.~\ref{fig:pk_comparison}), and the \galcatdesi case (red dashed line), corresponding to a stronger FOTO signal (see blue stars in Fig.~\ref{fig:pk_comparison}). While the former leads to negligible bias in the  $f_{\rm nl}$ measurement (where the bias is negative, consistent with the fact that for that realization $\Delta P_0<0$ at large scales, see Fig.~\ref{fig:FOTO_comp}), the latter induces a shift of $\Delta f_\mathrm{nl} \approx 24 = 1.5 \,\sigma_{f_{\rm nl}}$. We compare the measurements of each catalogue with their respective best-fit model in Fig.~\ref{fig:pk_comparison}. The lower panel shows the residuals between the data and the model in units of the standard deviation. In all cases, the data remain within $3\sigma$ over most of the $k$-range, except at the largest scales of the \galcatfsWr{\mc{O}} and \galcatdesiWr{\mc{O}_s}measurements, where the residuals are driven by the high FOTO signal.

\begin{figure}
        \centering
        \includegraphics[width=\linewidth]{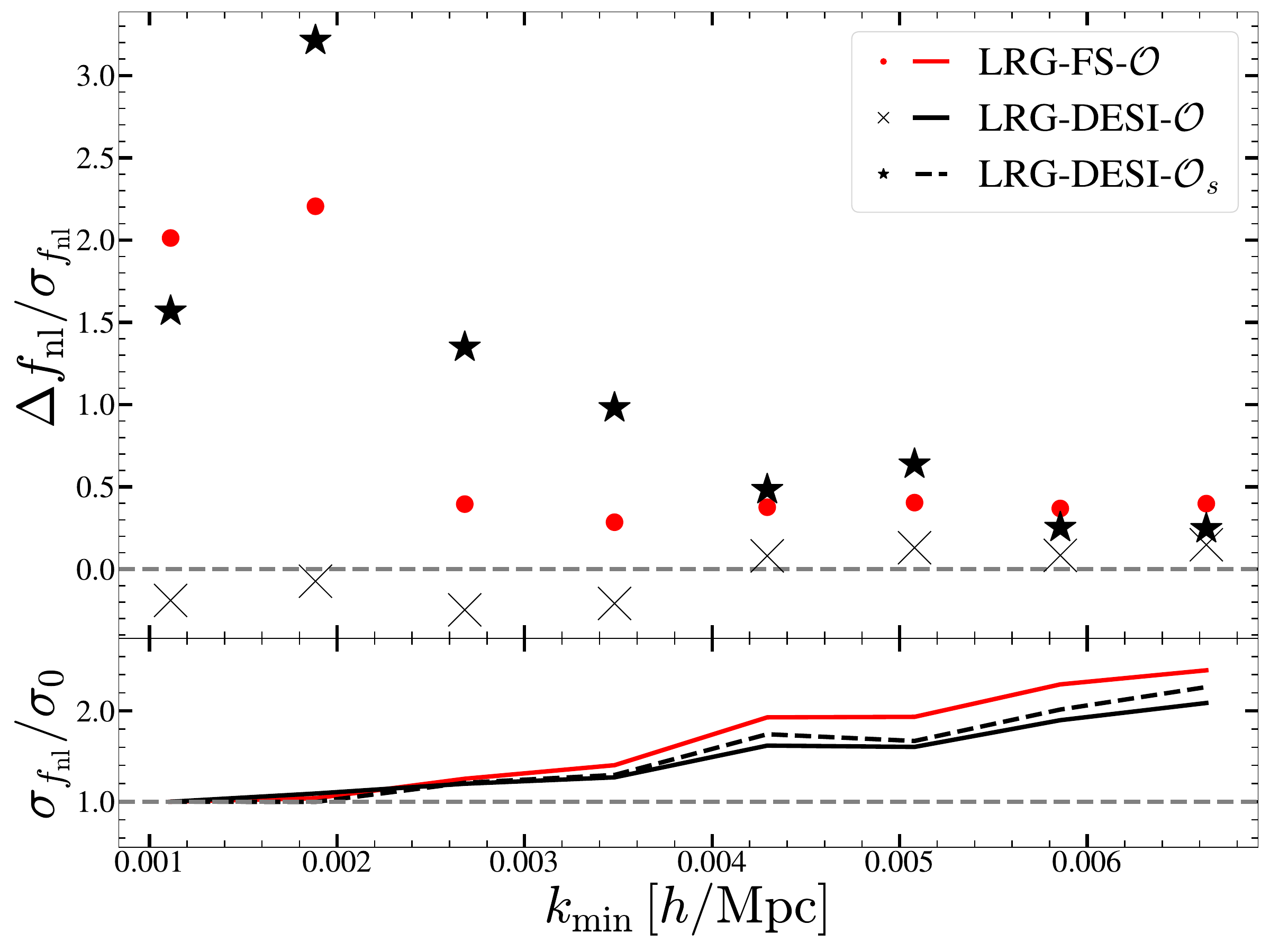}
        \caption{\justifying \textit{Upper:} Bias in the $f_\mathrm{nl}$ measurement due to the FOTO signal plotted as a function of the minimum $k$ employed in the analysis, for the full-sky sample \galcatfsWr{\mc{O}} (red dots), and the masked-sky samples \galcatdesiWr{\mc{O}} (black crosses) and \galcatdesiWr{\mc{O}_s} (black stars). The bias is shown in units of the standard deviation of the measurement, which differ between the DESI-like and full-sky samples. \textit{Lower:} Change in constraining power as a function of $k_\mathrm{min}$ presented as $\sigma_{f_\mathrm{nl}}/\sigma_{0}$, where $\sigma_{0}=\sigma_{f_\mathrm{nl}}(k_{\rm min}=10^{-3})$. The colour scheme matches the upper panel.}
        \label{fig:fnl_kmin}
        
\end{figure}

The amplitude of the FOTO signal is both oscillatory and rapidly  decreasing  with $k$. To assess the scale at which the bias induced by this signal is removed, we repeat the parameter inference while varying only the minimum wavenumber $k$, included in the analysis. We compute the quantity $\Delta f_\mathrm{nl} /\sigma_{f_\mathrm{nl}}$ for the different pairs shown in Fig.~\ref{fig:fnl_comparison} as a function of $k_{\rm min}$. The resulting biases are displayed in Fig.~\ref{fig:fnl_kmin} for \galcatfsWr{\mc{O}}  (red points), \galcatdesiWr{\mc{O}} (grey crosses), and \galcatdesiWr{\mc{O}_s} (black stars). We also present the relative constraining power of the measurements, quantified as $\sigma_{f_\mathrm{nl}} /\sigma_{0}$, shown with lines following the same colour scheme, where $\sigma_{0}=\sigma_{f_\mathrm{nl}}(k_{\rm min}=10^{-3})$   in the lower panel. For the \galcatfs case, the bias falls below $0.5\,\sigma_{f_\mathrm{nl}}$ for $k_{\rm min} > 2 \times 10^{-3}\,h/\mathrm{Mpc}$, while it remains negligible at all scales for the \galcatdesiWr{\mc{O}} realization. In contrast, for the partial-sky \galcatdesiWr{\mc{O}_s}  realization, the bias persists up to $k_{\rm min} \approx 4 \times 10^{-3},h/\mathrm{Mpc}$. We emphasize that this behaviour depends on the survey selection (see Eq.~\ref{alpha_0}) and the redshift bin considered; thus, for different selection, a dedicated study is required to establish the appropriate $k_{\rm min}$. However, increasing the $k_{\rm min}$ to avoid this bias entirely nearly doubles the size of the constraints on the $f_{\rm nl}$ as shown in the lower panel of Fig.~\ref{fig:fnl_kmin}. 

\begin{figure}
        \centering
        \includegraphics[width=\linewidth]{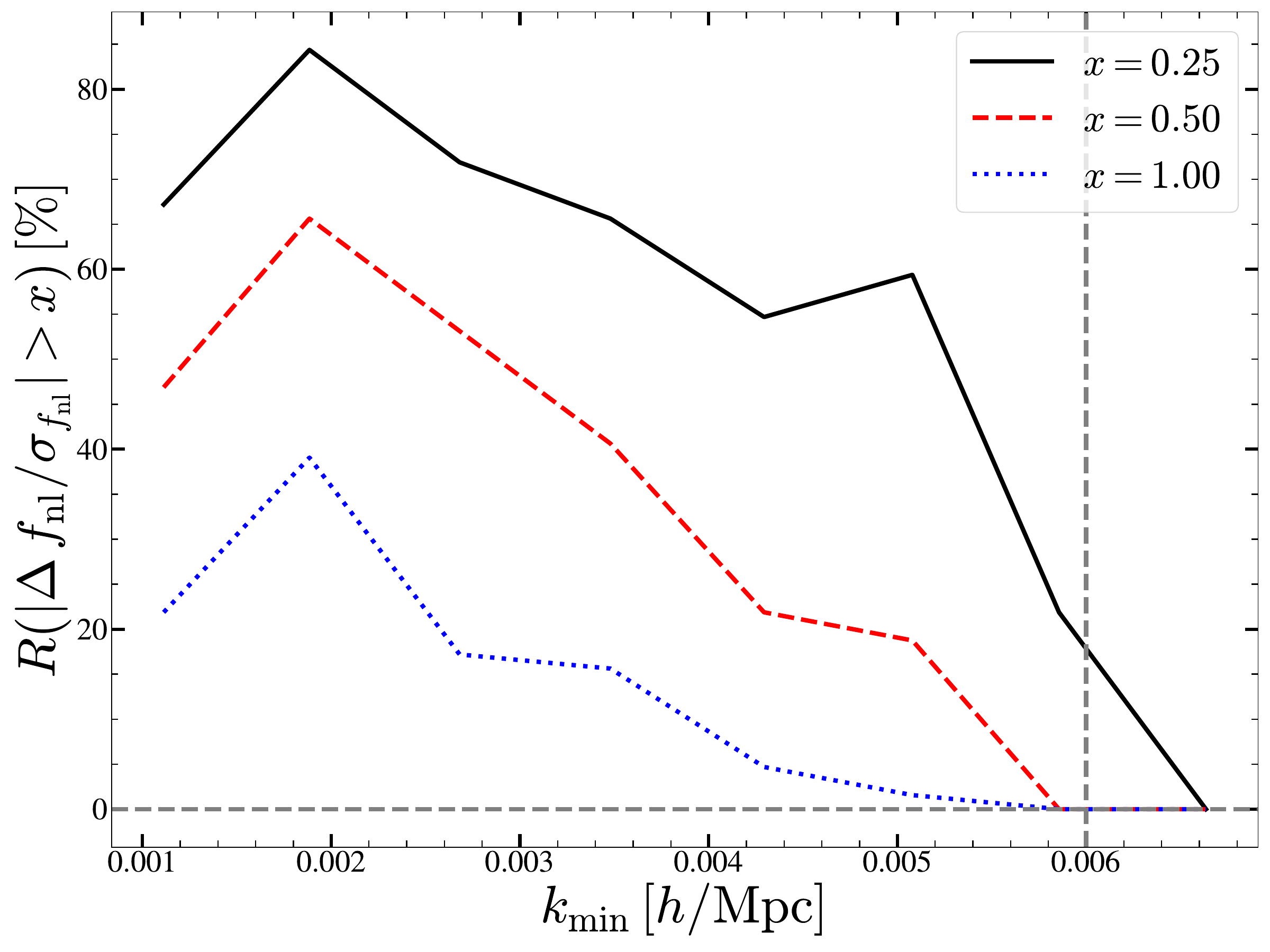}
\caption{\justifying Fraction of masked-sky \bcgalcatdesi mock realizations yielding a bias $|\Delta f_\mathrm{nl}/\sigma_{f_\mathrm{nl}}| > x$ for different thresholds $x$, as a function of the $k_\mathrm{min}$ adopted in the analysis. The grey vertical line shows the $k_\mathrm{min}$ used for the LRG $f_\mathrm{nl}$ constraints of \citet{chaussidon2024constrainingprimordialnongaussianitydesi}.}
    
    \label{fig:biases_k}
\end{figure}

As illustrated in Figs.~\ref{fig:FOTO_comp} and \ref{fig:fnl_kmin}, different realizations of the FOTO signal can produce significantly different biases in the $f_\mathrm{nl}$ measurements. To account for this, we repeat the analysis across all $64$ FOTO realizations, each time imprinting the signal onto the $\mathcal{G}$-sample power spectrum multipoles of the \galcatdesi measurements. In Fig.~\ref{fig:biases_k}, we present the fractional percentage of realizations, $R(x,k_\mathrm{min})$, that result in a bias in $f_\mathrm{nl}$ exceeding $x,\sigma_{f_\mathrm{nl}}$ as a function of $k_\mathrm{min}$. For $x=1$ (blue dotted line), we find that approximately $20\%$ of the realizations produce a bias larger than the measurement error when including scales below $0.003\,h/\mathrm{Mpc}$. As progressively larger modes are excluded, $R(x,k_\mathrm{min})$ converges to zero. Similar behaviour is observed for the $x=0.5$ (red dashed line) and $x=0.25$ (black solid line) cases, with higher values of $R(x,k_\mathrm{min})$ that decay less rapidly as $x$ decreases. In particular, for all values of $x$, the convergence of $R(x,k_\mathrm{min})$ towards zero is not monotonic. This is consistent with Fig.~\ref{fig:fnl_kmin}, where the oscillatory nature of the FOTO signal implies that including larger scales can, in some cases, slightly reduce the bias in $f_\mathrm{nl}$. 

Several remarks help contextualize the interpretation of the preceding results. Firstly, we ignore the impact of various angular systematics \citep[e.g., Milky Way extinction and variations of the photometric zero-point][]{Ross_2012,DES:2015vnr,Elsner:2015aga,euclidcollaboration2025euclidpreparationcontrollingangular}, which strongly affect power-spectrum measurements on large scales, as our goal is to demonstrate the significance of the FOTO signal even in the best-case scenario. Secondly, \citet{chaussidon2024constrainingprimordialnongaussianitydesi} found that the geometrical systematic mitigation techniques of the DESI LRG sample remained valid down to $k_\mathrm{min} = 10^{-3}\,h/\mathrm{Mpc}$, and the imaging systematics mitigation down to $k_\mathrm{min} = 3 \times 10^{-3}\,h/\mathrm{Mpc}$.
However, in their measurements they conservatively limited the analysis to $k_\mathrm{min} = 6 \times 10^{-3}\,h/\mathrm{Mpc}$. 
Since for this scale cut we find that there is a probability $\sim 20\%$ to present biases larger than $0.2 \,\sigma_{f_{\rm nl}}$, and taking into account that the DR1 sky coverage they probed is smaller than the Y5-like mask used in this work, we do not expect the FOTO to have substantially biased their LRG measurements. Thirdly, in this analysis, we choose to vary the type of mocks while keeping the model restricted to the peculiar velocity contributions, rather than using only the $\mc{O}$ mocks and explicitly adding the FOTO effect in the model. This approach is motivated by two considerations: (i) it allows us to explore different realizations of the FOTO signal, rather than only its expectation value, and (ii) accounting analytically for the survey geometry is complex \citep[see][]{Elkhashab_2025}, since the FOTO effect generates both odd and even multipoles that extend beyond $\ell = 4$. Nevertheless, for a Gaussian likelihood with the same covariance, the average shift in $f_{\rm nl}$ computed by varying the mocks is equivalent to the shift that would be induced by including the FOTO effect directly in the model. Finally, we note that the FOTO signal of our sample has an average signal-to-noise ratio (S/N) of around 1 for the partial-sky case. A higher S/N \citep[as shown for the Euclid Spectroscopic Sample, where $\mathrm{S/N}\approx4\,$,][] {Elkhashab_2021} would lead to stronger biases in the $f_{\rm nl}$ constraints.

\subsubsection{Relativistic effects impact on full-shape analysis}

In this subsection we examine the impact of relativistic RSDs on the cosmological parameters inferred from the full-shape analysis of the power spectrum. To that end, we infer the cosmological parameters from the \galcatdesi catalogue both with and without relativistic RSDs, directly measuring their impact parameter inference. For the theoretical modeling of the power spectrum, we adopt the effective field theory (EFT) of the large-scale structure, which describes the observables of the galaxy clustering through perturbative expansions of the nonlinear density and velocity fields \citep[for a recent review, see][]{ivanov2022effectivefieldtheorylarge}.
The EFT  model divides the theoretical power spectrum into three main contributions. The first contribution stems from the dynamical, bias, and RSD terms computed at the leading and next-to-leading order in standard perturbation theory (SPT). That contribution is parameterized by the cosmological parameters $\{h,\omega_{\rm c},\,\omega_{\rm b},\, A_{\rm s},\,n_{\rm s}\}$ and the bias parameters  $\{b_1,b_2,\gamma_2,\gamma_{21}\}$.  The second arises from the stochastic components of the density and velocity fields, represented by the stochastic shot-noise parameters: $\{N^p_0,N^p_{20},N^p_{22}\}$ . Whereas the previous contributions are already present in SPT, EFT introduces a set of parameters that encode small-scale corrections not captured by SPT. These corrections are implemented via “counterterms”,  ${c_0, c_2, c_4, c_{\rm nlo}}$.

We compute the model predictions via the Gaussian-process emulator \texttt{comet} \citep{10.1093/mnras/stac3667}, with priors summarized in Table~\ref{table:Priors_fit}. All parameters are varied except $\gamma_{21}$ and $N^p_{22}$, with $\gamma_{21}$ set by the coevolution relation
\begin{equation}
\gamma_{21} = \frac{2}{21}(b_1 -1 ) + \frac{6}{7}\gamma_2\,,
\end{equation}
and $N^p_{22}$ fixed to zero. We perform parameter inference  performed using the  importance nested sampling algorithm implemented in \texttt{Nautilus}, and the chains are analyzed and visualized with \texttt{getdist}.

\begin{table}
\caption{\justifying \label{table:Priors_fit} The prior distributions used for parameter inference in
  the EFT model. Cosmological parameters are separated by a horizontal line from
  the bias and nuisance parameters. The symbols $\mc{N}$ and $\mc{U}$ represent
  Gaussian and uniform distributions, respectively.}
\def\arraystretch{1.12}
\begin{center}
\begin{tabular}{lr}
Parameter & Prior  \\
\hline
$h$              &$\mc{U} (0.5,1.0)$\\
$\omega_{\rm c}$  &$\mc{U} (0.085, 0.155)$\\
$A_{\rm s}$       &$\mc{U} (1.4, 2.6)$\\
$n_{\rm s}$       &$\mc{N} (0.96, 0.0041)$ \\
$\omega_{\rm b}$  &$\mc{N} (0.02218, 0.00055)$\\
\\
$b_1$            & $\mc{U}(0.25,4.0)$ \\
$b_2$            & $\mc{U}(-5,5)$\\
$\gamma_2$       & $\mc{U}(-5,5)$\\
$\gamma_{21}$    & Coevolution \\
$c_0\,[\mpc^2]$              &$\mc{N}(0,50)$\\
$c_2\,[\mpc^2]$              &$\mc{N}(0,50)$ \\
$c_4\,[\mpc^2]$              &$\mc{N}(0,50)$\\
$c_{\rm nlo}\,[\mpc^4]$      &$\mc{N}(0,50)$\\
$N^P_0$              & $\mc{N}(0,3)$ \\
$N^P_{20}\,[\mpc^2]$              & $\mc{N}(0,50)$ \\

\end{tabular}
\end{center}
\end{table}

\begin{figure}
    \centering
    \includegraphics[width=1\linewidth]{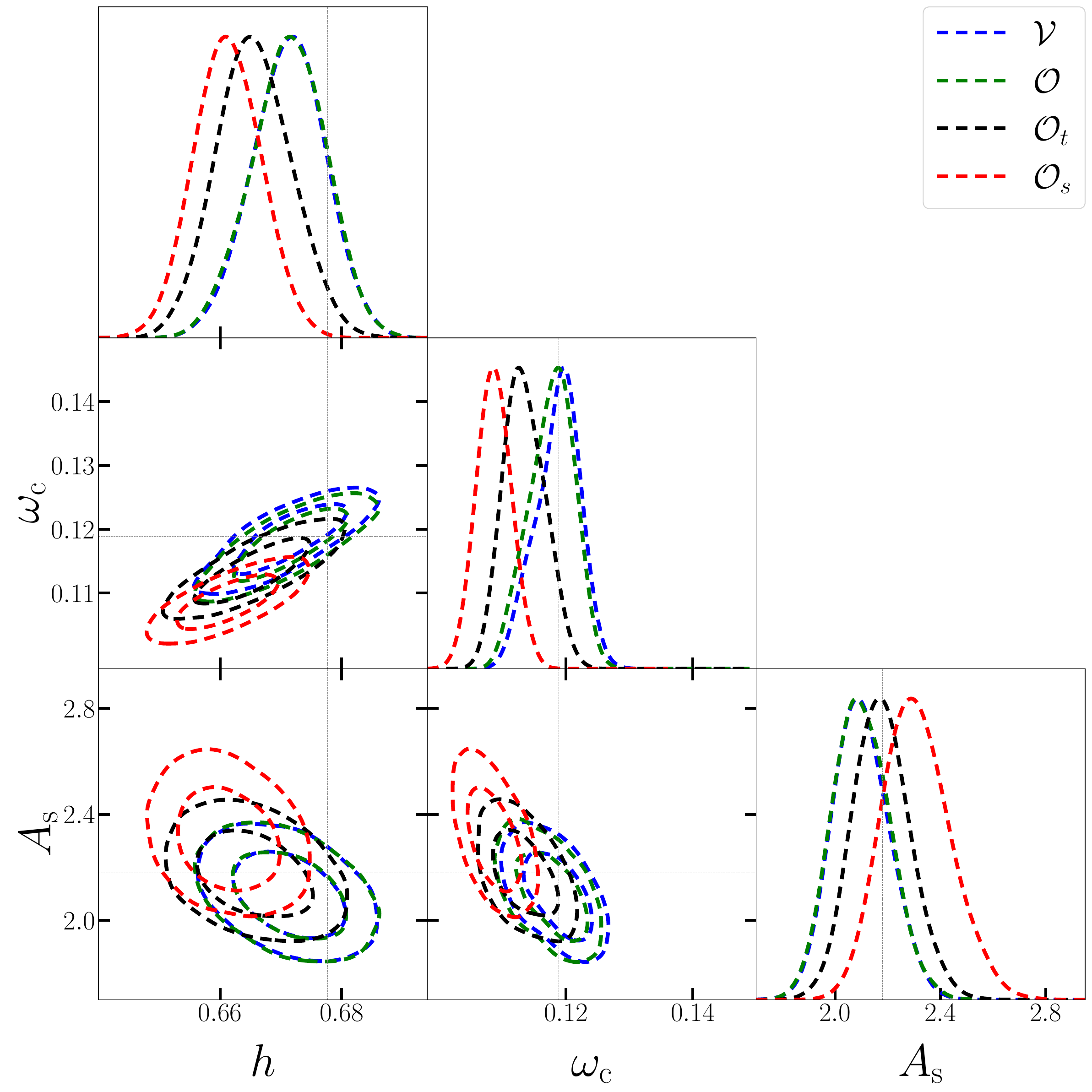}
    \caption{\justifying Corner plot showcasing shifts in cosmological constraints due to different relativistic effects from the \galcatdesi catalogues.  }
    \label{fig:Corner_Full_shape}
\end{figure}

We perform parameter inference using the monopole and quadrupole of the \galcatdesi power spectrum over the range $k \in (2 \times 10^{-3}, 0.2)\,h/\mathrm{Mpc}$, with bin width $\dif k = 3 \times 10^{-3}\,h/\mathrm{Mpc}$. We employ the $\mc{V}$, and $\mc{O}$ catalogues to measure the impact of all  relativistic RSDs, along with two variations of the $\mc{O}$ catalogue in which we imprint a typical ($\mc{O}_{\rm t}$) and a strong ($\mc{O}_{\rm s}$) FOTO signal (dotted and solid black lines in  Fig.~\ref{fig:Corner_Full_shape}). We show the inferred cosmological parameters in Fig.~\ref{fig:Corner_Full_shape}. We can deduce from the figure that adding relativistic RSDs with a weak FOTO signal has little impact on the parameter constraints, as we can see from the overlapping contours of the $\mc{V}$ and $\mc{O}$ mocks.
The typical and strong FOTO signal, on the other hand, induce a systematic shift in all three parameters shown. That shift increases in amplitude as the FOTO signal strength increases. We note that the posterior means are consistent with the maximum-likelihood estimate derived from the MCMC chain at the few-percent level, suggesting that prior-volume effects are negligible \citep[e.g.][]{Carrilho_2023,Simon_2023}.
To identify the $k$-range at which this bias is significant, we use the Figure-of-Bias (FoB).

We define the FoB as 
\begin{equation}
    {\rm FoB} = \left[(\bs{\theta}_{a}-\bs{\theta}_{\mc{V}})\,S_{a}^{-1}\, (\bs{\theta}_{a}-\bs{\theta}_{\mc{V}})^T\right]^{1/2}\,,
\end{equation}
where, $\bs{{\theta_{a}}}$ are vectors containing the
posterior means of the three cosmological parameters, the subscript denotes the type of catalogue $a\in\{\mc{O},\mc{O}_{\rm t},\mc{O}_{\rm s}\}$, and $S_{a}$ is
the parameter covariance matrix, computed from each MCMC chain using the \texttt{getdist} package. Assuming that  the covariance does not change between the different chains  and that the posterior is a Gaussian distribution, the FoB quantifies the covariance-weighted difference between the posterior means of the mocks with relativistic RSDs and their velocity-only counterpart.

\begin{figure}
    \centering
    \includegraphics[width=1\linewidth]{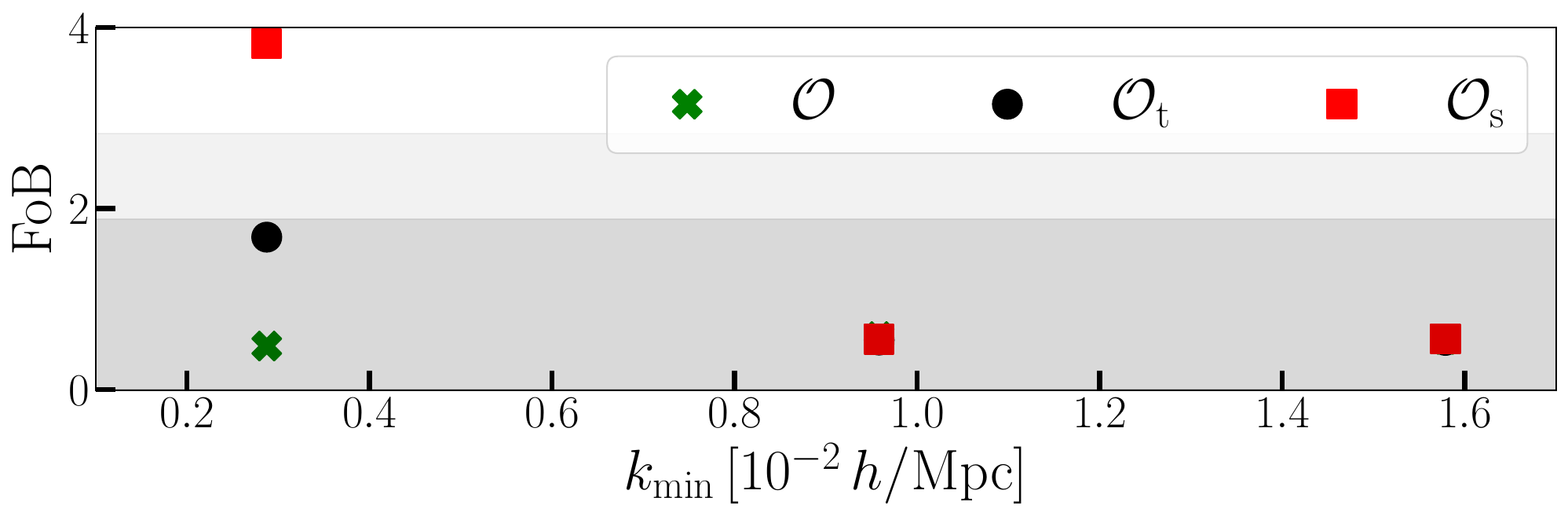}
    \caption{\justifying FoB values obtained from different catalogues w.r.t. the $\mc{V}$ mocks are plotted against the minimum scale included in the parameter inference.}
    \label{fig:GOODNESS}
\end{figure}

The covariance matrix converts the
parameter differences into levels of significance, such that a value of $\mathrm{FoB}=
1.88\,(2.83)$ can be interpreted as corresponding to the $68\%\,(95\%)$ credibility levels.
We recompute the chains varying the minimum scales considered and plot the FoB in Fig.~\ref{fig:GOODNESS}. As seen from the figure, the FOTO signal leads to a 4$\sigma$ bias at the largest scales considered for the \galcatdesiWr{\mc{O}_s} mocks. However, that bias is removed by excluding the largest scales from parameter inference, showing that as we consider the typical range of scales for a full-shape analysis, the impact of the FOTO is negligible.


\section{Conclusions} 
\label{sec:conc}
Stage-IV galaxy surveys will measure redshifts and angular positions for tens of millions of galaxies over a large fraction of the sky, spanning redshift ranges that allow us to probe clustering on previously inaccessible scales. In constructing galaxy maps from these surveys, it is standard to assume that galaxies populate an unperturbed background universe when converting redshifts into comoving distances. However, galaxy positions and fluxes are influenced by the inhomogeneities encountered by the light along its path from source to observer, leading to various RSDs. While the leading RSD effect, arising from peculiar velocity differences between sources and the observer \citet{1987MNRAS.227....1K}, is routinely accounted for, additional relativistic contributions—including Doppler terms, gravitational lensing, Sachs–Wolfe effects, and time delays—have often been neglected. These corrections become increasingly relevant on scales approaching the Hubble radius. In this work, we present a new implementation of the \liger method, \ligerGAL, that incorporates relativistic effects directly at the tracer level in mock halo/galaxy catalogues derived from high-resolution Newtonian N-body simulations. Our method makes it possible to post-process high-resolution simulations by imprinting relativistic RSDs by using the halo merger tree to produce a halo catalogue with the effects included.

We also developed a toolkit that populates the generated halo catalogue with galaxies or other tracers, consistently with the relativistic RSDs imprinted by \liger. For a chosen halo population model (such as HOD) the toolkit generates galaxy positions that incorporate the full relativistic treatment of super-halo scales, while accounting for peculiar velocity distortions of individual galaxies, on sub-halo scales. These contributions affect not only their line-of-sight positions but also the Doppler-induced contribution to magnification bias, as described in equations \ref{v_shift_gal} and \ref{v_mag_gal}.

We designed this approach to align with the formalism of the \texttt{HaloTools} library, ensuring straightforward adaptability to a variety of halo population models \footnote{Available at:
\url{https://gitlab.com/liger-cosmo/liger-hod}}.

We applied our method to the HMDPL simulation, constructing a full-sky halo catalogue that reaches $z=0.8$. We compared the \ligerGAL implementation to the previous field-based one (\ligerDM), and showed that we can extend the accuracy of the produced catalogues to capture the non-linearities of the halo clustering signal, where \ligerDM fails. 
Having access to the non-linear scales allowed us to perform realistic measurements of clustering observables, such as the galaxy power spectrum, which we in turn used to perform standard cosmological inference analyses, to assess the impact of large-scale relativistic effects in the inference process. 

Then, relying on the HOD formalism, we generated a DESI-like LRG sample that presents the imprint of relativistic RSDs at large scales. We utilised a full-sky and a partial-sky sample of that catalogue incorporated with different levels of relativistic RSDs to assess the impact of relativistic contributions which are generally not accounted for in the power spectrum models used in the standard full-shape analysis, as well as PNG inference via galaxy clustering.

In particular, we assessed the impact of the observer’s peculiar motion (the FOTO effect) on the inference of the $f_\mathrm{nl}$ parameter. By focusing directly on the bias $\Delta f_\mathrm{nl}$ caused by the FOTO, we were able to disentangle our results from other possible effects that could impact the analysis, such as the radial integral constraint \cite{de_Mattia_2019}, or exploring different values of the \textit{true} underlying $f_\mathrm{nl}$. We found that this contribution leads to a bias in the measurement of $f_\mathrm{nl}$ both for the full- and partial-sky surveys, which varies with the largest scale included in the analysis. 
In particular, we found that for a DESI-like (partial sky) analysis, the FOTO signal biases $f_\mathrm{nl}$ by more than $1\sigma_{f_\mathrm{nl}}$ in approximately 40\% of the realizations of the universe when including scales down to $k_\mathrm{min}=1.5\times 10^{-3}\,h/\mathrm{Mpc}$; this probability drops to ~10\% if scales below $k_\mathrm{min}<5\times 10^{-3}\,h/\mathrm{Mpc}$ are excluded. 
Full-shape analyses of the power spectrum monopole and quadrupole using the EFT model indicate that $A_{\rm s}$, $\omega_{\rm c}$, and $h$ are biased when including scales below $k_{\rm min} < 4 \times 10^{-3}\,h/\mathrm{Mpc}$, but this bias disappears if these large scales are excluded.  Due to the size of available simulations, our analysis was limited to an LRG-like sample up to $z=0.8$. Extending to higher-redshift populations requires larger simulations capable of producing wide light-cone catalogues. We note that the FOTO signal can vary substantially with survey selection, and consequently, the biases discussed here may be amplified for different populations or redshift ranges. Nonetheless, the toolkit introduced here can be applied directly to generate light-cones with relativistic RSDs, provided sufficiently large simulations with halo merger trees are available. 

Since upcoming measurements of $f_\mathrm{nl}$ will probe larger regions of the Universe and include increasingly larger scales (e.g.\ DESI Y5, Euclid DR3), the FOTO contribution will become progressively more significant. It is therefore important to incorporate this signal into both the power-spectrum model and the covariance matrices, along with other wide-angle systematics such as Milky Way extinction, and large-scale flux variations. Incorporating these effects into mock catalogues is relatively straightforward, allowing for a  robust estimation of their impact on parameter constraints. 
Applying the same treatment in analytic models is more challenging, as it requires careful accounting of the mixing between scales and multipoles—including odd multipoles—induced by the survey geometry. Nevertheless, analytic approaches to the FOTO signal could offer the additional advantage of simultaneously constraining the observer's velocity and PNG, or of marginalizing over the velocity parameters. We leave the development of such strategies to future work.

\begin{acknowledgements}
We thank Benedict Bahr-Kalus, Pierluigi Monaco, Emiliano Sefusatti and Licia Verde for the useful suggestions and discussion during the first distribution of this manuscript.
DB acknowledges support from the COSMOS network (www.cosmosnet.it) through ASI (Italian Space Agency) Grants 2016-24-H.0, 2016-24-H.1-2018 and 2020-9-HH.0
We acknowledge the use of computational resources from the
University of Padova Strategic Research Infrastructure Grant 2017: “CAPRI: Calcolo ad Alte Prestazioni per la Ricerca e l’Innovazione”. 
This paper is supported by the PRIN 2022 PNRR project "Space-based cosmology with Euclid: the role of High-Performance Computing" (code no. P202259YAF), funded by European Union – Next Generation EU.
We acknowledge the Gauss Centre for Supercomputing e.V. (www.gauss-centre.eu) and the Partnership for Advanced Supercomputing in Europe (PRACE, www.prace-ri.eu) for funding the MultiDark simulation project by providing computing time on the GCS Supercomputer SuperMUC at Leibniz Supercomputing Centre (LRZ, www.lrz.de). The Bolshoi simulations have been performed within the Bolshoi project of the University of California High-Performance AstroComputing Center (UC-HiPACC) and were run at the NASA Ames Research Center.
\end{acknowledgements}
\bibliographystyle{aa} 
\bibliography{aanda} 

\begin{appendix}
\section{Lightcone construction}
In this appendix, we outline the procedure for constructing the halo lightcone and computing the gravitational potential at the relevant redshifts for the \ligerGAL method.

\subsection{Halo trajectories}
\label{lightcone}

To construct the observed past light cone from the simulation, we implement a cubic spline interpolation procedure for halo positions as they intersect with the observer's lightcone. 
We first select an observer position at the $z=0$ snapshot and assign it a peculiar velocity; if none is specified, the code adopts the velocity of the nearest halo to ensure consistency with the local environment. 
For each halo, we reconstruct its full trajectory across all simulation timesteps, starting from the earliest snapshots until a merger occurs or the trajectory is interrupted. 
Subsequently, we identify the instant at which the halo intersects the observer's past light cone and apply the interpolation scheme of \cite{Elkhashab_2021}, while additionally tracking the halo mass evolution between snapshots.

Regarding the definition of halo trajectories, several caveats are considered to ensure a consistent catalogue.  
When a merger between two halos occurs at snapshot $i$ (where $i=0$ corresponds to $z=0$ and $z_i > z_{i-1}$) we treat the Most Massive Progenitor (MMP) and the secondary merging halos differently. For the MMP, we continue to follow the evolution of its descendant in snapshot $i-1$. For non-MMP halos, we add a final step to their trajectory in snapshot $i-1$, keeping their mass fixed but assigning them the position and velocity of the descendant. 
Although the halo merger occurred in between the two snapshot, it is not possible to determine if it occurred before or after the lightcone intersection. In this work, we assume the former case, so that the two haloes intersect the lightcone as a single halo.

\begin{figure}
\centering
    \includegraphics[width=\linewidth]{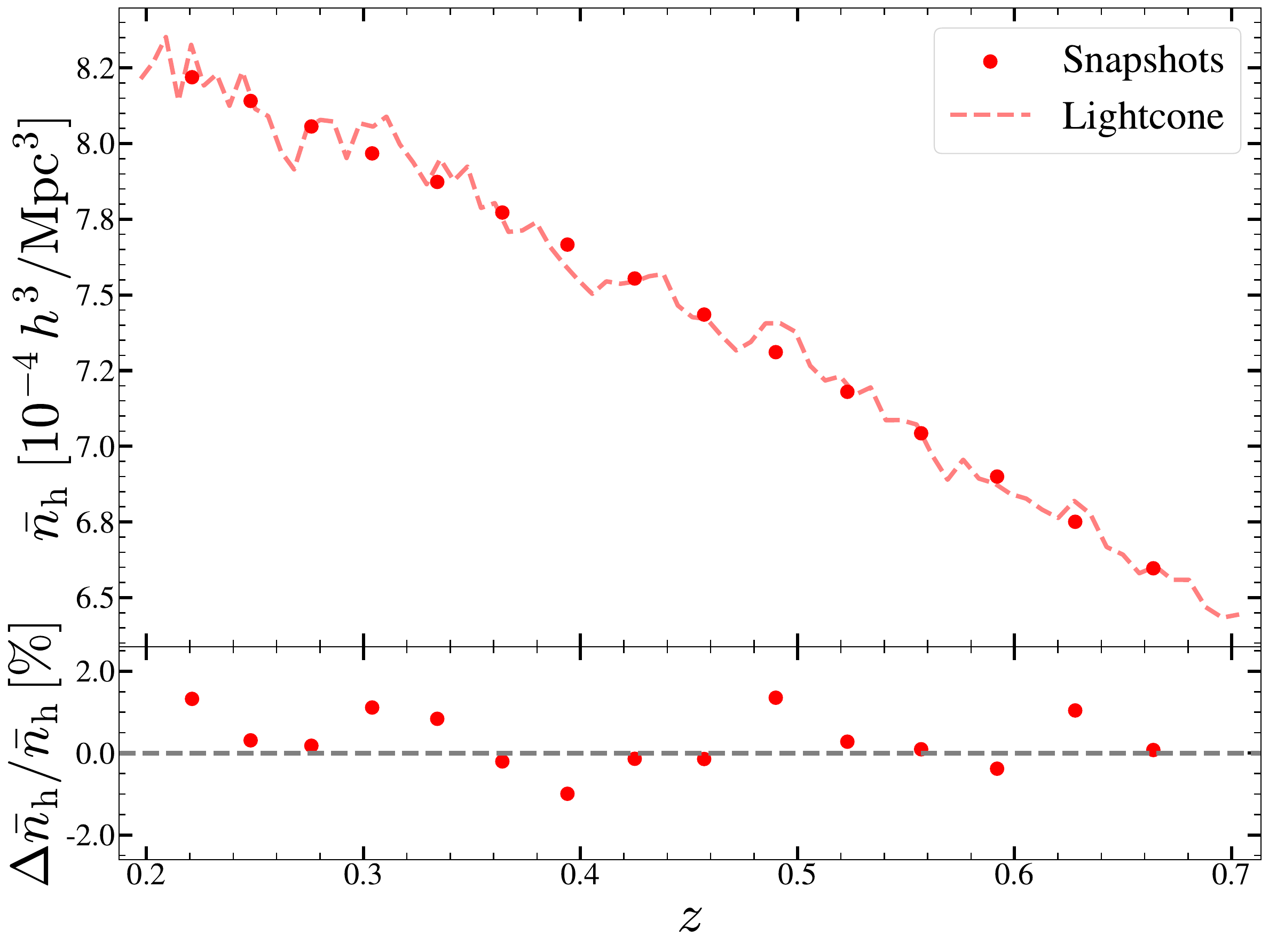}
\caption{\justifying In the upper panel, we compare the halo radial number density of the \dmcatfs{1} within the redshift bin $z\in[0.2,0.7]$ (red dashed curve) to the values computed at each simulation snapshot (dotted markers). The lower panel shows the relative difference between these quantities.}
\label{fig:radial_test}
\end{figure}

Another situation of interest occurs when halos without progenitors are first identified in a given snapshot $i$ and do not intersect the light cone in subsequent snapshots. In such cases, no information is available on whether they may have crossed the light cone between snapshots $i$ and $i+1$ ($z_{i}<z_{i+1}$). This can happen either because the halo actually formed in snapshot $i$, or because the halo finder failed to correctly link it to its progenitor in snapshot $i+1$. To address this issue, we initially excluded these halos; however, this approach systematically underestimated the radial density of halos with $M_\mathrm{vir}\geq 5\times 10^{12},\mathrm{M}_\odot/h$ compared to snapshot-based estimates.
The actual importance of this effect depends on the merger-tree definition of the simulation under consideration and may be less impactful in other simulations. Consequently, we opt to always account for the possibility of a light-cone intersection between $i$ and $i+1$. We do so by extrapolating the halo position at the snapshot $i+1$ by assuming a motion with  constant velocity $\bs{\varv}_\textit{i}$, taken to be the halo's velocity at the snapshot \textit{i}. Under this prescription, the extrapolated position assigned to these halos is  $\bs{x}_{\textit{i}+1}=\bs{x}_\textit{i}+\bs{\varv}_\textit{i}\cdot \left(t_\textit{i+1}-t_\textit{i}\right)\,,$ where $t_i$ is the proper time of the snapshot. To validate the latter approach, we compare the radial densities of the \dmcatfs{1} with the radial densities obtained from the halo snapshots. 
The results are shown in Fig. \ref{fig:radial_test}. Each red dot corresponds to the radial density of a snapshot halo catalogue at the corresponding radial distance, and they agree (within a few percent) with the dashed curve, which represents the radial density of the \dmcatfs{1} catalogue.

\begin{figure}
\centering
    \includegraphics[width=\linewidth]{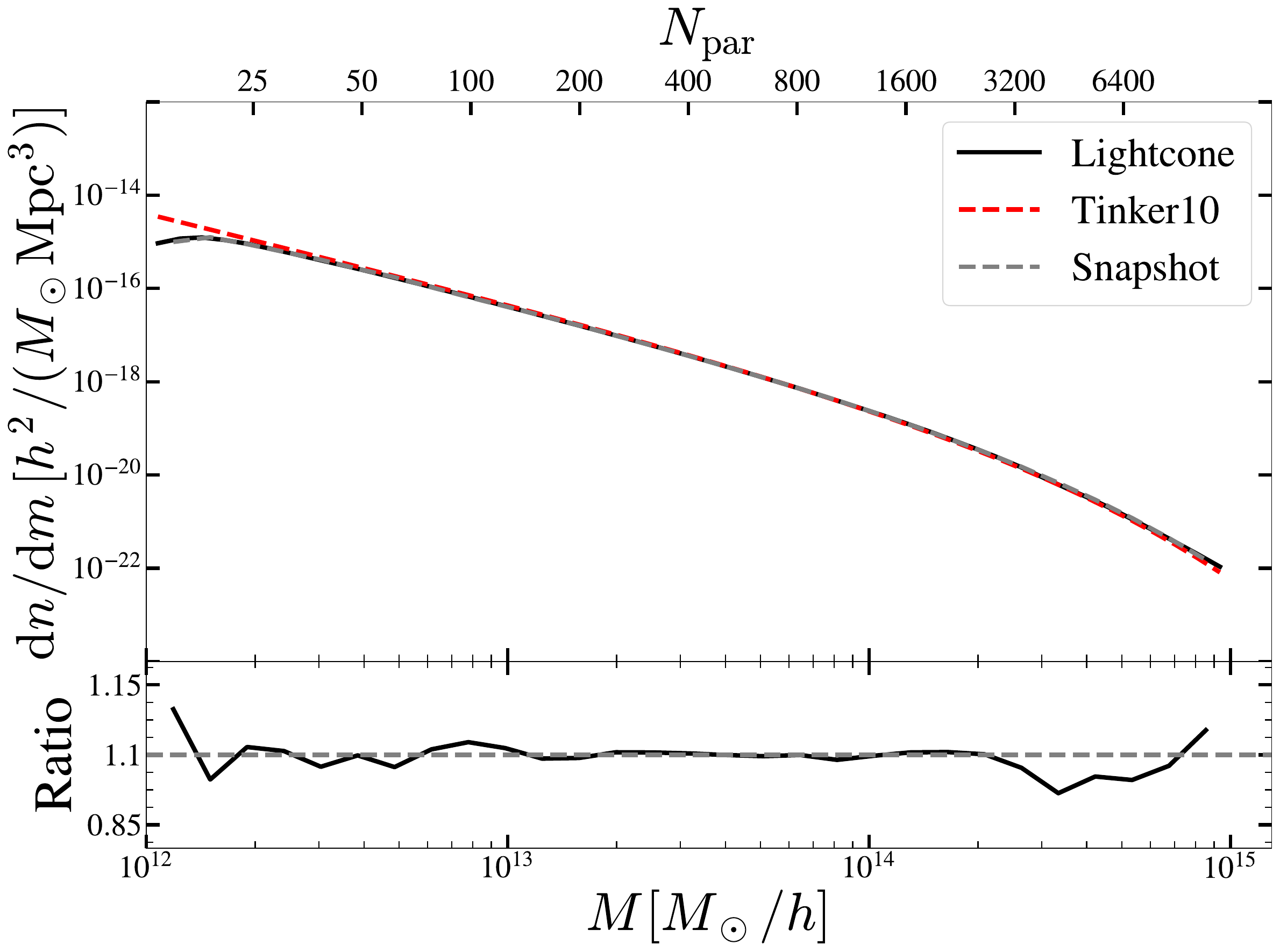}
\caption{\justifying In the upper panel, we compare the HMF of  \dmcatfs{1} lightcone (black solid)  computed within a narrow redshift bin centred at $z=0.46$ with thickness $\Delta z = 0.06$ to the snapshot HMF at $z=0.46$ (grey dashed). We also plot the corresponding  Tinker10 model  prediction (red dashed line). In the bottom, we show ratio of the between the lightcone and snapshot HMFs.}

\label{fig:hmf}
\end{figure}

Finally, we validate our halo mass interpolation procedure in Fig. \ref{fig:hmf}, where we compare the halo mass function (HMF) obtained from a thin shell of the \dmcatfs{1} (black solid line) with the HMF estimated from a snapshot of the simulation at the mean redshift of the shell (gray dashed line). For comparison, we also plot the HMF model of \citet{Tinker_2010} (hereafter Tinker10) evaluated at the snapshot redshift. As evident in the figure, the \galcatfs{1} HMF agrees with the snapshot HMF across the entire mass range. Both HMFs deviate from Tinker10 at the low-mass end due to the mass resolution limit of the HMDPL simulation.

\subsection{Gravitational potential computation}

\label{potential}
The underlying principle of the \liger method is that, in a $\Lambda$CDM cosmology at linear order, the Bardeen potentials (see Eq.~\ref{eq:shift_lig}) coincide, i.e. $\Phi = \Psi$, and can be computed by solving the Poisson equation, with the matter overdensity from the input simulations serving as the source term \citep[see][]{Borzyszkowski:2017ayl,Elkhashab_2025}.
As the HMDPL simulation provides only two publicly available particle snapshots, we used linear theory to model the time evolution of the gravitational potential, allowing us to derive the potential at the redshifts of the available halo snapshots. The procedure is as follows. First, we computed the potential at one of the matter snapshots using spectral methods implemented with the FFTW library \citep{1386650}. The potential can then be obtained at any scale factor $a$ in a given grid cell located at position $\boldsymbol{x}_i$, we use the linear relation
\begin{equation}
\label{pot_ev}
\Phi(\boldsymbol{x}_i, a) = \frac{a_0\,D^{(+)}(a)}{a\,D^{(+)}(a_0)} \, \Phi(\boldsymbol{x}_i\, a_0)\,,
\end{equation}
where $D^{(+)}$ is the linear growth factor of density perturbations and $a_0$ denotes the reference scale factor of one of the available snapshots.

We test the validity of the linear approximation using the simulation data by computing the gravitational potential from the two available particle snapshots on a grid with $N_{\rm grid} = 256^3$, and comparing the average of their ratio with the theoretical prediction from Eq.~(\ref{pot_ev}). The mean ratio, estimated over all grid cells where neither of the two potentials vanishes, is
\begin{equation}
\hat{r}\equiv\left\langle \frac{\Phi(z_1)}{\Phi(z_2)}\right\rangle\bigg\vert_{\Phi(z_1),\Phi(z_2)\neq 0}=1.15496\,,
\end{equation}
while the theoretical prediction reads
\begin{equation}
    r= \frac{D^{(+)}(a_1)\,a_2}{D^{(+)}(a_2)\, a_1}=1.15477\,,
\end{equation}
that implies a sub-percent fractional error of  $(\hat{r}-r)/{r}=0.016\%\,$.

\begin{figure}
\centering
    \includegraphics[width=\linewidth]{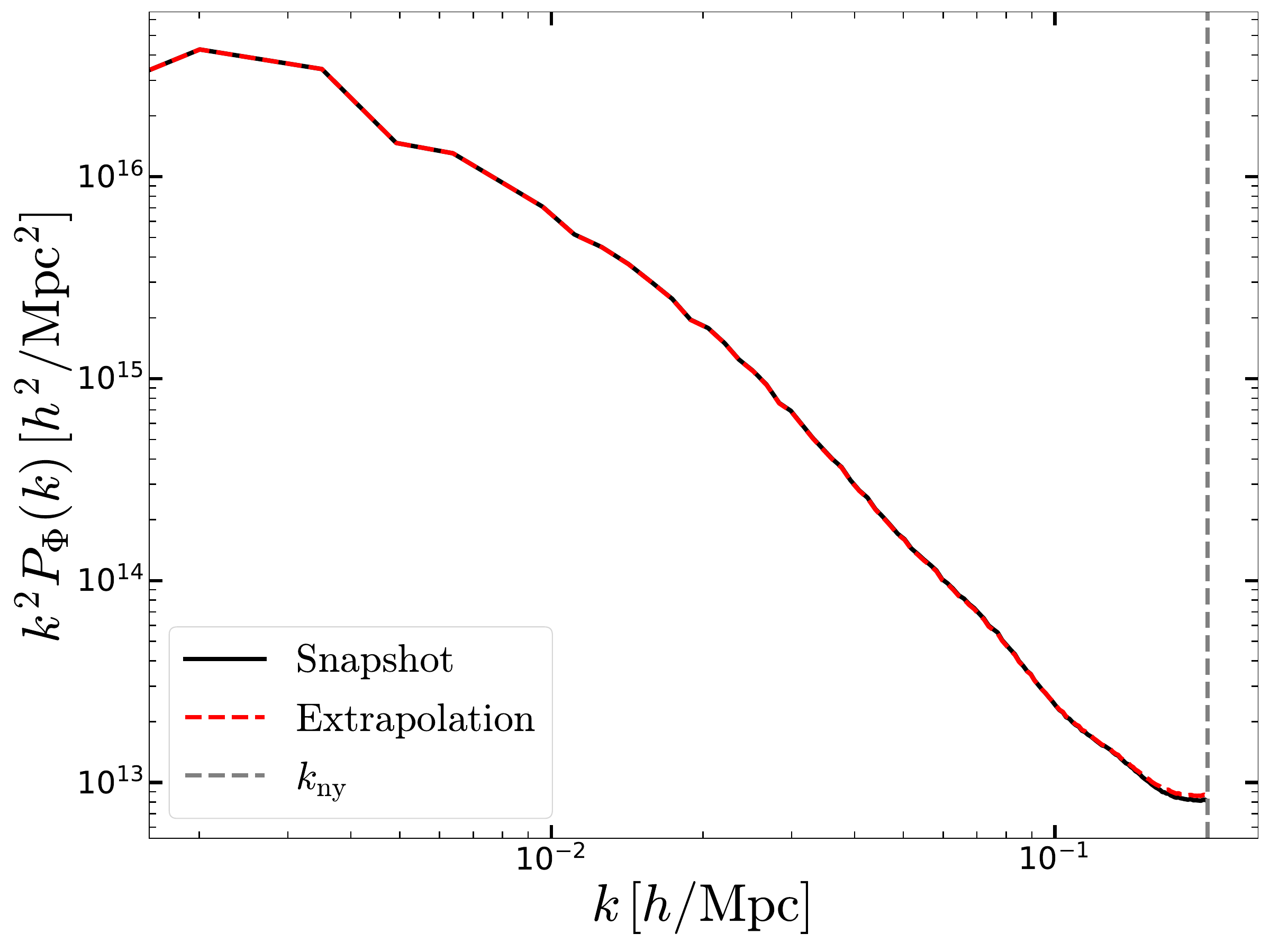}
\caption{\justifying Power spectrum comparison of the gravitational potential $\Phi$ obtained respectively from the particle distribution at the HMDPL snapshot $z=0.49$ (black-continuous line) and from the linear theory extrapolation of equation \ref{pot_ev}, starting from the $z=0.0$ snapshot (red-dashed line). In grey we show the Nyquist frequency of the measurement.}
\label{fig:potential_test}
\end{figure}

We also test whether the linear solution accurately recovers the potential power spectrum, $\bs{P}_\Phi(k)$, at different redshifts. In Fig.~\ref{fig:potential_test}, we compare the power spectrum of the gravitational potential computed directly from the snapshot at redshift $z_2$ (black solid line) with the linear extrapolation (red dashed line) obtained using Eq.~\ref{pot_ev}. The two estimates show excellent agreement across all scales up to the Nyquist frequency of the grid, $k_\mathrm{Ny} \approx 0.20\,h\,\mathrm{Mpc}^{-1}$. We therefore conclude that the linear approximation is sufficient for our purposes.
Consequently, we extrapolate the potential to all redshifts by utilizing the potential at $z=0$ and applying Eq.~\ref{pot_ev}. Finally, we note that this assumption is not strict: the potential extrapolation is not required for the implementation of our approach and can be relaxed when full particle snapshots (or sufficiently large subsamples) are available at each redshift.

\begin{figure}
\centering
    \includegraphics[width=\linewidth]{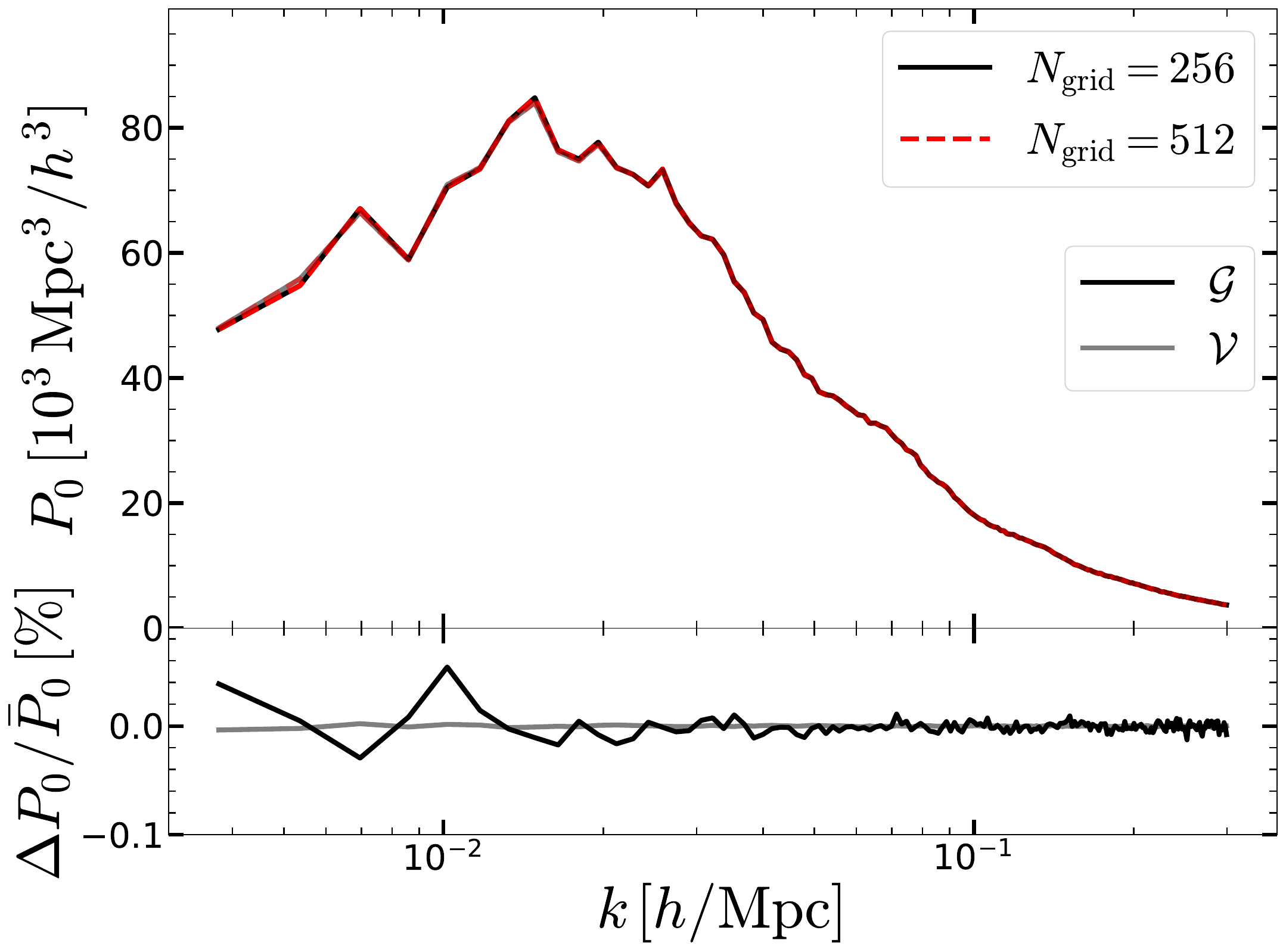}
\caption{\justifying In the upper panel we show the power spectrum monopole of the \dmcatfs{} catalogue in the $z\in[0.6-0.8]$ bin, computed at two different potential grid resolutions (labelled with black-full and red-dashed lines), for both the $\mathcal{V}$ and $\mathcal{G}$ mocks, showed respectively with black and grey lines). In the lower panel we show the residuals between the two resolutions for each mock.}
\label{fig:res_test}
\end{figure}

Last, we assess whether the resolution of the potential might affect the clustering signal at scales under study in this work. In the upper panel of Figure \ref{fig:res_test} we show the power spectrum monopole of the \dmcatfs{} catalogue in the $z\in[0.6,0.8]$ bin, computed for the $\mathcal{G}$ (opaque lines) and $\mathcal{V}$ (semi-transparent lines) mocks, at two different resolutions of the gravitational potential: $N_\mathrm{grid}=256$ (black-full lines) and $N_\mathrm{grid}=512$ (red-dashed lines). In the bottom plot we show relative difference between the two potential resolutions for each sample: at all the scales probed, the impact of a coarser gravitational potential is below $0.1\%$ for the halo clustering signal. Thus, in order to be consistent with our linear treatment of the potential evolution, we choose to use $N_\mathrm{grid}=256$.

\subsection{Galaxy gravitational terms approximation}
\label{gal_approx}
In this section, we show how the approximation of the gravitational terms made in section \ref{galaxy_sample} holds when probing scales much larger than the typical halo size.
Let the centre of a halo be located at
\begin{equation}
\mathbf{x}_\mathrm{h} = x_\mathrm{h} \hat{\mathbf n}_\mathrm{h}\, ,
\end{equation}
and the position of a galaxy inside the halo be
\begin{equation}
\mathbf{x}_\mathrm{g} = \mathbf{x}_\mathrm{h} + \mathbf X_\mathrm{g}\,,
\qquad
|\mathbf X_\mathrm{g}|\lesssim R\, ,
\end{equation}
where $R$ denotes the typical halo size. Decomposing the displacement as
\begin{equation}
\mathbf X_\mathrm{g} = X_\mathrm{g}^\parallel\, \hat{\mathbf n}_\mathrm{h} + \mathbf X_\mathrm{g}^\perp\, ,
\end{equation}
for any scalar field $F(\mathbf x)$, a Taylor expansion around the halo position gives
\begin{equation}
\delta F
\equiv F(\mathbf{x}_{\mathrm{g}})-F(\mathbf{x}_{\mathrm{h}})
=
X_\mathrm{g}^\parallel\,\partial_\parallel F
+
\mathbf X_\mathrm{g}^\perp\cdot\nabla_\perp F
+
\mathcal{O}\!\left(X_\mathrm{g}^2\,\nabla^2F\right),
\end{equation}
where $\partial_\parallel$ and $\nabla_\perp$ are parallel and transverse components  of the gradient w.r.t. $\hat{\mathbf{n}}_\mathrm{h}$, and the $\mathcal{O}(X_\mathrm{g}^2\nabla^2 F)$ term collects higher-order contributions involving higher powers of $X_\mathrm{g}$ and higher derivatives of $F$.

This expansion can be applied to both local and integrated contributions of the gravitational potential of  the galaxy overdensity (see Eq.\ref{eq:Deltag}), such as lensing convergence or gravitational redshift. 
For a Fourier mode $\mathbf k = k_\parallel\, \hat{\mathbf n}+ \mathbf k_\perp$,
\begin{equation}
\partial_\parallel \sim k_\parallel\,,
\qquad
\nabla_\perp \sim k_\perp\,,
\end{equation}
so the leading correction scales as
\begin{equation}
\delta I \sim (kR)\, I\,.
\end{equation}
Hence, on scales larger than the halo size ($kR\ll1$), galaxy--halo displacements
produce only suppressed corrections.

A further suppression, however, arises once the galaxy distribution inside each halo is taken into account. Let $u_\mathrm{h}(r)$ denote the normalised intra-halo galaxy profile, 
\begin{equation} 
\int d^3r\,u_\mathrm{h}(r)=1\,,
\end{equation} 
which we assume here to be isotropic, consistently with the HOD model implemented.
We can then write the Fourier transform of $F(\mathbf{x})$ as 
\begin{equation} 
F(\mathbf{k}) = \sum_\mathrm{h} e^{-i\mathbf{k}\cdot\mathbf{X}_\mathrm{h}} \sum_{a\in \mathrm{h}} e^{-i\mathbf{k}\cdot\mathbf{X}_{\mathrm{a},\mathrm{h}}} \,F(\mathbf{x}_\mathrm{h} + \mathbf X_{\mathrm{g},\mathrm{a}})\,,
\end{equation} 
where the index ``$\mathrm{h}$" spans over the halos of the sample, and ``$\mathrm{a}$" over the galaxies belonging to each halo. Replacing the discrete galaxy distribution within each halo by the profile $u_\mathrm{h}(\mathbf r)$ gives 
\begin{equation} 
F(\mathbf{k}) = \sum_{\mathrm{h}} e^{-i\mathbf{k}\cdot\mathbf{X}_{\mathrm{h}}} \int d^3r\,u_{\mathrm{h}}(\mathbf r) \,e^{-i\mathbf{k}\cdot\mathbf r}\, F(\mathbf{X}_{\mathrm{h}}+\mathbf r)\,. 
\end{equation} 
If we expand the argument of the integral over $r$, we find that the first-order terms are dipolar, and  thus average out by the integration with the halo profile. The  leading order corrections are  a combination of quadrupolar terms
\begin{equation} 
\begin{split}
\delta F(\mathbf{k})\approx \sum_{\mathrm{h}} e^{-i\mathbf{k}\cdot\mathbf{X}_{\mathrm{h}}}[A k_j \partial_i I(\mathbf{X}_{\mathrm{h}})\langle r_i r_j\rangle_{\mathrm{h}}+B\partial_i\partial_j I(\mathbf{X}_{\mathrm{h}})\langle r_i r_j\rangle_{\mathrm{h}}] \,.
\end{split}
\end{equation} 
Then, since $\partial_i\sim k_i$, $\partial_i\partial_j\sim k_i k_j$ and $\langle r_i r_j\rangle_{\mathrm{h}}\sim R^2$, the correction scales as 
\begin{equation} 
\delta F \sim (kR)^2 F\,,
\end{equation}
which leads to a stronger suppression than in the previous result.

\section{Survey functions numerical estimation}
\label{survey_est}

In this appendix, we briefly describe the procedure implemented to estimate the survey functions of the \dmcatfs{1}, \galcatfs, and  \galcatdesi. These survey functions [precisely $\bar{n}(z)$, $b_1(z)$, $\mc{E}(z)$ and $\mc{Q}(z)$]. In particular, for the \dmcatfs{1} catalogue, we compute the survey functions from the simulation snapshots, while for the \galcatfs and \galcatdesi sample we work directly with the lightcone data.

\subsection{\dmcatfs{1} Catalogue}

As the halos are available in snapshot format,  we can compute the survey functions directly at each redshift from the snapshots. We compute the $\bar{n}(z_i)$ (where $i$ runs over the different snapshots) simply by counting the haloes belonging to the sample and dividing by the simulation comoving volume. For the linear bias, we compute from the power spectra ratio. For a given snapshot at a redshift $z_i$, and a sample of mass-selected halos, we use the linearly evolved   matter power spectrum $P_m(k_j,z_i)$ (see appendix ~\ref{potential}) and estimate  the halo power spectrum $P_{m}(k_j,z_i)$, where $k_j$ are the discrete wavenumbers at which the power spectrum is sampled.
Then,  we estimate the linear halo bias using an average of the power spectra ratio
\begin{equation}
\label{bias_est}
    b_1(z_i)= {\frac{1}{N_k} }\sum_{k_j}^{k_j<k_0}\sqrt{\frac{P_{\rm h}(k_j,z_i)}{{P}_{\rm m}(k_j,z_i)}}
\end{equation}
where $k_0$ is an upper limit for $k$ that we choose so that the linear bias relation holds and $N_k = \sum_{k_j}^{k_j<k_0} $.

For the magnification and evolution bias, we use \citep[see appendix B in ][]{Elkhashab_2021}
\begin{equation}
\label{eq:dlnNdlnz}
    -\frac{\mathrm{d}\,\mathrm{ln}\,\bar{n}(z)}{\mathrm{d}\,\mathrm{ln}\,(1+z)}=2\mathcal{Q}(z)\,\left[1+\frac{(1+z)}{H(z)\,x(z)}\right]+\mathcal{E}(z)\,.
\end{equation}

As for the halo sample $Q(z) = 0$ by construction, 
we directly note that 
\begin{equation}
    \mathcal{E}(z) = -\frac{\mathrm{d}\,\mathrm{ln}\,\bar{n}(z)}{\mathrm{d}\,\mathrm{ln}\,(1+z)}\,.
\end{equation}

\subsection{\galcatfs catalogue}
\label{LRG_bias}
We use an analogous approach to construct the survey functions for the \galcatfs and \galcatdesi catalogues, albeit applied directly on the lightcone, since the galaxies are painted directly onto the lightcone halo catalogue. To compute $\bar{n}$, we count the galaxies in redshift bins of size $\delta z = 0.04$, and divide by the comoving volume associated with each corresponding spherical shell.

For the linear bias, we use the angular power spectra ratio. Specifically, we compute the ratio between the angular power spectra of the \galcatfsWr{\mc{R}} catalogues in tomographic redshift bins, $C_\ell(z_i)$ (where $z_i$ is the median redshift of the bin), and a theoretical linear angular power spectrum computed using \ttt{CAMB}. The linear bias is then obtained by averaging over a range of multipoles:
\begin{equation}
b_{1}(z_i) = {\frac{1}{N_\ell}} \sum_\ell^{\ell < \ell_0} \sqrt{\frac{C_\ell^{\mc{R}}}{C_\ell^{\mathrm{Linear}}}}\,,
\end{equation}
where $\ell_0$ is the maximum multipole chosen such that the linear biasing relation holds, and $N_\ell = \sum_\ell^{\ell < \ell_0} $.

For the magnification bias, we use the estimates from \citet{Zhou_2023}, where the complex selection function of the LRGs is already taken into account. We interpolate their measurements to each galaxy’s redshift.

There is, however, a slight subtlety in the computation of the evolution bias. Since the LRG galaxies are selected using flux cuts across several bands, we account for magnification effects through a weighting scheme (see Eq.~\ref{wmag}). This implies that the logarithmic derivative in Eq.~\ref{eq:dlnNdlnz} must be applied to $\bar{n}_{\mathrm{g}}=\bar{w}_{\mathrm{g}}(z)\tilde{n}_{\mathrm{g}}(z)$, where $\tilde{n}_{\mathrm{g}}(z)$ denotes the average density of the catalogues without the magnification weights, and $\bar{w}_{\mathrm{g}}(z)$ is the average value of the weight $w_{\mathrm{g}}(\bs{x})$ at a given redshift.
Consequently, we estimate $\mathcal{E}(z)$ from the logarithmic  total derivative of the galaxy number density \textit{before} applying the magnification weights.

\section{Estimators}
\label{estimators}
In this appendix, we review the estimators utilised the two-point statistics studied in this work. 

\subsection{Angular power spectrum}
\label{cl_estimator}

For the estimation of the  angular power spectrum, we make use of the HEALPix algorithm \citep[Hierarchical Equal Area isoLatitude Pixelisation][]{Gorski_2005} to partition the sky into $N_\mathrm{pix} = 12 \times N_\mathrm{side}^2$ pixels.
For each tracer sample $i$, we compute the projected counts within a redshift bin, $N^{i}_{\rm g}(\Omega)$, and then estimate the projected density contrast as
\begin{equation}
\Sigma^{i}(\Omega) = \frac{N^i_\mathrm{g}(\Omega)}{\bar{N}^{i}_\mathrm{g}}-1\,,
\end{equation}
where $\bar{N}^i_{\mathrm{g}}$ is the average of the projected counts over the survey footprint.
Then, exploiting the fast spherical harmonic decomposition of HEALPix, we compute
\begin{equation}
    a^i_{\ell,m}=\int\Sigma^{i}(\Omega)\,Y^*_{\ell m}(\Omega)\,\dif^2\Omega\,,
\end{equation}
which we use to estimate the auto and cross angular power spectra with the pseudo-$C_\ell$ estimator \citep[][]{1973ApJ...185..413P}
\begin{equation}
    \bs{C}_l^{ij} = \frac{1}{w_\mathrm{p}^2(2l+1)f_\mathrm{sky}}\sum_{m=-l}^l a_{lm}^i {a_{lm}^j}^*-\frac{\delta_K^{i,j}}{\bar{N}_\mathrm{g}^i}\,,
\end{equation}
where $f_\mathrm{sky}$ is the sky fraction covered by the survey footprint, and $w_\mathrm{p}$ is a pixelisation correction factor.

\subsection{3D power spectrum}
\label{pk_theory}
For the (3D) power spectrum multipoles, we first construct the FKP field \citep[][]{Feldman_1994}:
\begin{equation}
\label{fkp}
    F(\bs{x})=\frac{w(\bs{x})}{\sqrt{A}}\left[{n}_\mathrm{g}(\bs{x})-\alpha\,{n}_\mathrm{r}(\bs{x})\right]\,,
\end{equation}
where $A=\int w^2(\bs{x})\, \bar{n}^2 \,\dif^3 x$ is a normalization factor, and the fields  $\bs{n}_{\rm g}\,\&\,\bs{n}_{\rm r}$ denote the number density of the data and random catalogues, respectively.  Moreover, \begin{equation}
    \alpha = \frac{\int w(\bs{x})\,n_{\rm g}(\bs{x}) \dif ^3x }{\left[\int w(\bs{x})\,n_{\rm r}(\bs{x}) \dif ^3x\right]}
\end{equation}  rescales the random density to match the data catalogue number densities. Finally, the FKP weights $w$ are given by
\begin{equation}
\label{w_fkp}
    w(\bs{x}) = \mathcal{I}(\bs{x})\left[1+\bar{n}(x)\mathcal{P}_0\right]^{-1}\,,
\end{equation}
where $\bar{n}(x)$ is the mean density of tracers in our catalogue at a distance $x$, $\mathcal{I}$ is the indicator function, equal to $1$ inside the survey volume and to $0$ elsewhere, and the parameter $\mathcal{P}_0$ represents an approximate value of the power spectrum at the scales under study.  In this work, we estimate $\bar{n}(x)$ from our random catalogue, evaluating it within radial bins  of thickness $ \delta r = 15\,\mathrm{Mpc}/h$ and interpolating it with a cubic spline.
For each of our samples, we generate random catalogues $50\times$ as dense as our data ($\alpha\approx 1/50$), sampling their distribution using the shuffling method, which assigns randomly selected redshifts from the original dataset and uniformly distributed angular positions within the survey mask. For the typical value of the power spectrum, we adopt $\mathcal{P}_0=10\,000\,\mathrm{Mpc}^3\,h^{-3}$.

To compute the power spectrum, we use the \texttt{pypower} code, employed in recent DESI data analyses. This implementation follows the FFT-based Yamamoto–Bianchi formalism \citep{yamamoto_cosmological_1999, Scoccimarro_2015, Bianchi:2015oia}:
\begin{equation}
    {P}_\ell(k)=(2\ell+1)\int F_\ell(\bs{k})F_0(-\bs{k})\frac{\mathrm{d}^2\Omega_k}{4\pi}-{P}^{\mathrm{SN}}_\ell(k)\,,
\end{equation}
where 
\begin{equation}
    F_\ell(\bs{k}) = \int F(\bs{x})\,e^{-{\rm i} \bs{k}\cdot\bs{x}} \mathcal{L}_\ell (\bshat{k}\cdot\bshat{x})\, \mathrm{d}^3x\,,
\end{equation} 
and 
\begin{equation}
\label{shot-noise}
    {P}^\mathrm{SN}_{\ell}(k) = \frac{(1+\alpha)}{A}\int w^2(\bs{x})\,{n}_\mathrm{g}(\bs{x})\,\mathcal{L}(\bshat{k}\cdot\bshat{x})\,\dif^3 x\,.
\end{equation}

Lastly, for all likelihood evaluations in the various parameter inference tests, we estimate the covariance matrices of the power spectra following the analytical approach of \citet{Wadekar:2019rdu}. In particular, using the \texttt{THECOV} code \citep{Alves2024prep}, we compute the covariance matrices for our power spectrum measurements, including the effects of the survey geometry. For simplicity, we only consider the Gaussian component of the covariance. We compute the covariance matrices for both the \galcatfs and \galcatdesi catalogues. We note that these covariance matrices do not account for the imprint of relativistic RSDs present in our dataset as the aim of our analysis is to investigate the biases that arise when neglecting such effects.

\end{appendix}
\end{document}